\documentclass[pdflatex,sn-mathphys-num]{sn-jnl}

\usepackage{amsmath}
\usepackage{amssymb}
\usepackage{amsthm}
\usepackage{bbm}
\usepackage{color}
\usepackage{float}
\usepackage[T1]{fontenc}
\usepackage{graphicx}
\usepackage{hyperref}
\usepackage[utf8]{inputenc}
\usepackage{physics}
\usepackage{times}
\usepackage{txfonts}
\usepackage{xcolor}

\newlength{\gwidth} 
\newlength{\twodpwidth} 
\newlength{\threedpwidth} 
\hypersetup{
  colorlinks=true,linkcolor=blue,citecolor=blue,
  filecolor=blue,urlcolor=blue,breaklinks=true
}

\begin{document}

\title[Article Title]
{Scattering entropies of quantum graphs with several channels}


\author[1]{\fnm{Alison A.} \sur{Silva}\orcid{0000-0003-3552-8780}}
\email{alisonantunessilva@gmail.com}

\author*[1,2]{\fnm{Fabiano M.} \sur{Andrade}\orcid{0000-0001-5383-6168}}
\email{fmandrade@uepg.br}

\author[3]{\fnm{Dionisio} \sur{Bazeia}\orcid{0000-0003-1335-3705}}
\email{bazeia@fisica.ufpb.br}

\affil[1]{
  \orgdiv{Programa de P\'os-Gradua\c{c}\~ao em Ci\^encias/F\'{i}sica},
  \orgname{Universidade Estadual de Ponta Grossa},
  \orgaddress{
  \city{Ponta Grossa},
  \postcode{84030-900},
  \state{Paran\'a},
  \country{Brazil}}}

\affil[2]{
  \orgdiv{Departamento de Matem\'atica e Estat\'istica},
  \orgname{Universidade Estadual de Ponta Grossa},
  \orgaddress{
    \city{Ponta Grossa},
    \postcode{84030-900},
    \state{Paran\'a},
    \country{Brazil}}}

\affil[3]{
  \orgdiv{Departamento de F\'isica},
  \orgname{Universidade Federal da Para\'iba},
  \orgaddress{
  \city{Jo\~{a}o Pessoa},
  \postcode{58051-900},
  \state{Para\'iba},
  \country{Brazil}}}

\abstract{
This work deals with the scattering entropy of quantum graphs in many
different circumstances.
We first consider the case of the Shannon entropy and then the R\'enyi
and Tsallis entropies, which are more adequate to study distinct
quantitative behavior such as entanglement and nonextensive behavior,
respectively.
We describe many results associated with different types of quantum
graphs in the presence of several vertices, edges, and leads.
In particular, we think the results may be used as quantifiers
in models related to the transport in quantum graphs.\\

\noindent DOI:
\href{https://doi.org/10.1140/epjp/s13360-024-05460-0}
{10.1140/epjp/s13360-024-05460-0}

}


\maketitle

\section{Introduction}\label{sec1}

Entropies are in general used to work as quantifiers of diversity,
uncertainty, or randomness of a given system.
In this sense, investigations utilizing the concept of entropy have been
considered in many different areas of nonlinear science.
Among the several distinct possibilities, in this work, we will consider
entropies introduced by Shannon, R\'enyi and Tsallis related to quantum
graphs.
Quantum graphs are nicely studied in Ref. \cite{Book.Berkolaiko.2012},
and the associated entropies will be introduced below, connecting them
with statistical information.

As one knows, the Shannon entropy \cite{BSTJ.27.379.1948} is directly
connected to information in general, and it can be seen as a way to
express probability.
The R\'enyi entropy, on the other hand, can be seen as an index of
diversity and in this sense be directly related to ecology, economy and
statistics, for instance, where it intends to provide a quantitative
measure to account for the many different types of species or objects
there are in a community or dataset.
It is also connected with entanglement measures in a quantum system, and
in this case, it has been recently used in many different contexts in
high energy physics; see, e.g.,
Refs. \cite{Book.Renyi.1970,PRL.116.251602.2016,PRL.126.160601.2021,
PRL.126.141602.2021}
and references therein, and the results on charged R\'enyi entropies \cite{JHEP.2013.59.2013,PRL.129.021601.2022}.
The Tsallis entropy \cite{JSP.52.479.1988,Book.Tsallis.2009} is of
interest to investigate systems where nonextensive or nonadditive behavior plays an important role. The concept has been used to investigate physical behavior in several distinct areas; see, e.g. \cite{PRD.91.114027.2015} where hadron
spectra can be adequately described by  nonextensive statistical
mechanical distribution and also, the recent works on black hole entropy
\cite{PRD.104.084030.2021} and nonlinear wave propagation in
relativistic quark-gluon plasma \cite{EPJC.82.189.2022}.

In the present work, we shall further explore the use of the Shannon
entropy to quantify scattering in simple quantum graphs, as we have
studied before in \cite{PRA.103.062208.2021,PE.141.115217.2022}.
Here, however, we shall expand our investigation into two distinct
directions: first, we study the Shannon entropy associated to
scattering in simple quantum graphs in terms of the energy of the
incoming signal, and not only in the form of the average quantity that
we have introduced before.
The second line of investigation concerns the extension of the concept
of the Shannon entropy to the cases of the R\'enyi and Tsallis
entropies, which are supposed to be more appropriate to investigate
systems with distinct statistical contents, such as entanglement
measures and nonextensive behavior.
Since we do not know in principle how a quantum graph arrangement may
respond under the  Shannon, R\'enyi or Tsallis entropies, we think it is
worth studying these  possibilities, to see if we can quantify the
different behavior they certainly engender.

The study of the Shannon entropy intended to introduce another
quantifier of information in simple quantum graphs was first presented
in Refs. \cite{PRA.103.062208.2021,PE.141.115217.2022}.
It was inspired by the possibility to add a global quantifier associated
with a scattering process in a quantum graph.
In other areas of research, it was also motivated by several interesting
results, in particular, on the possibility to make  the energy flux
travel from a cold system to a hot system, by increasing  the Shannon
entropy of the memory to compensate for the decrease in the
thermodynamic entropy \cite{PRL.111.030602.2013}, and on the theoretical
framework for the thermodynamics of information, connected to stochastic
thermodynamics and fluctuation theorems, opening a way to manipulate
information at the molecular or nanometric scales \cite{NP.11.131.2015}.
Moreover, in high-energy physics, the concept of configurational
entropy, based on the Shannon entropy, was first introduced in
\cite{PLB.713.304.2012, PRD.86.045004.2012}.
It has also been used in some interesting directions to examine
stability of physical systems such as in the investigation of glueballs
\cite{PLB.765.81.2017}, heavy mesons \cite{PLB.776.78.2018} and
skyrmions \cite{JMMM.475.734.2019,PLA.392.127170.2021}, the
configurational entropy in AdS/QCD models \cite{PLB.820.136485.2021,PRD.103.106027.2021},
entanglement generation in meson-meson scattering process \cite{PRD.104.114501.2021} ,
the nuclear information entropy
\cite{EPJP.137.590.2022,EPJP.137.762.2022}, and in other
scenarios of recent interest
\cite{PRD.105.064049.2022,arXiv:2207.06367}.

\begin{figure}[t]
  \centering
  \includegraphics[width=\gwidth]{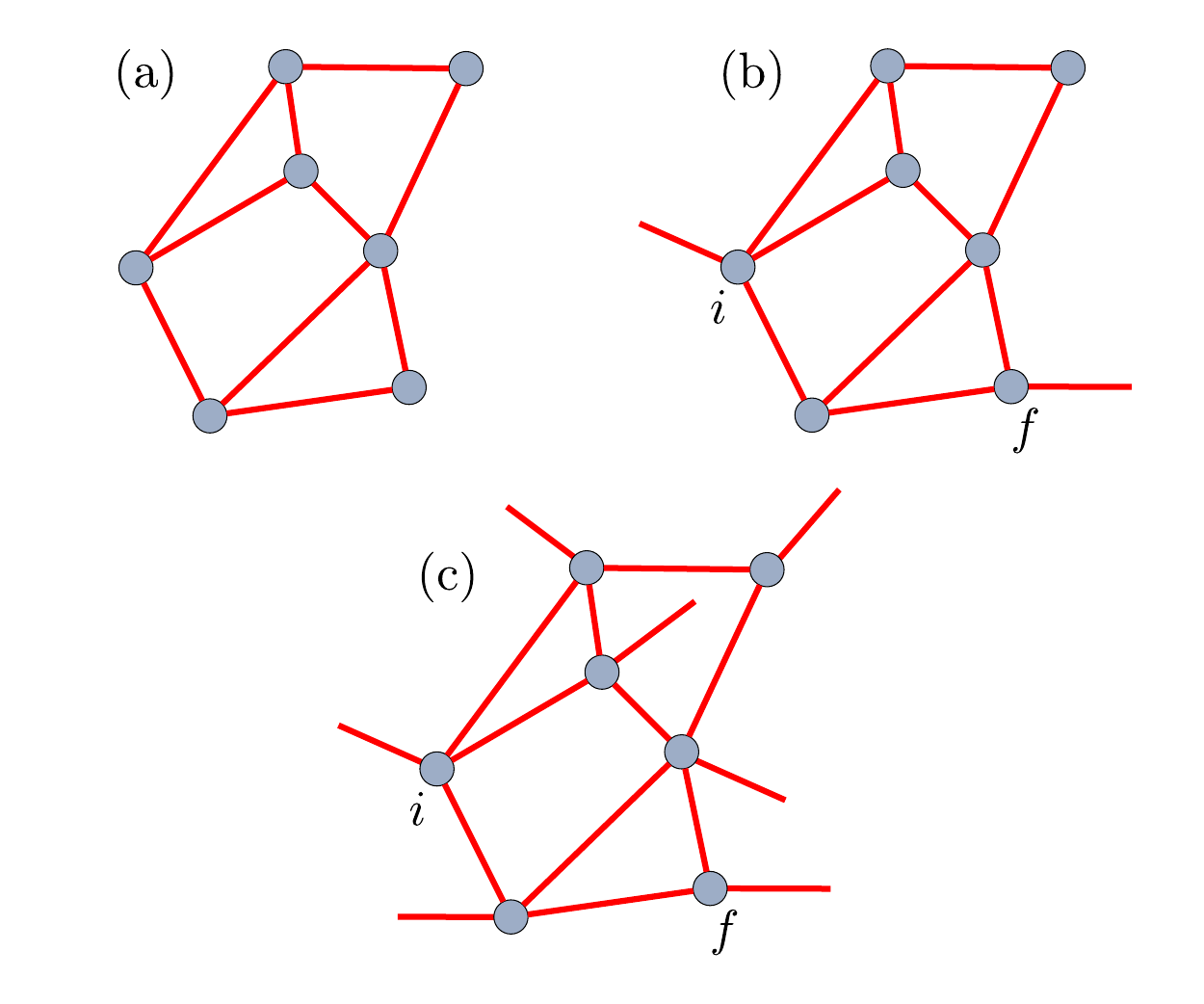}
  \caption{
    (a) Closed quantum graph with $7$ vertices and $10$ edges.
    (b) The associated open quantum graph with $2$ and
    (c) with $7$ leads added, with $i$ and $f$ identifying
    the entrance and exit scattering channels.
  }
  \label{fig:fig1}
\end{figure}

To describe the present investigation pedagogically, we organize the
work as follows.
In Sec. 2 we review two important issues, one being the scattering in
quantum graphs, and the other the concept of Shannon entropy associated
to the quantum scattering, and then we add the extensions to the case of
R\'enyi and Tsallis entropies.
To investigate the scattering one needs to make the quantum graph open,
and we do this by attaching leads to vertices in the graph, as it is
illustrated in the case of a graph with 7 vertices and 10 edges in
Fig. \ref{fig:fig1}.
In the same Sec. 2, we implement the main novelties of the work, which
concerns the calculations related to the Shannon, R\'enyi and Tsallis
entropies and we do this for both the scattering entropy, which depend
on the energy or the wave number of the incoming signal, and the average
scattering entropies, which are global quantities, averaged over the
period of the periodic scattering probabilities.
We then move on to illustrate the results with several distinct
examples, for the Shannon, R\'enyi and Tsallis entropies.
This is implemented in Sec. 3, where we compute the results for series
and parallel disposition of vertices, and also for the following types of
graphs: cycle, wheel and complete  graphs.
We compare some results, to show how these different quantities behave
as we vary the energy or wave number of the incoming signal in the
quantum graph.

\section{Scattering and entropies in quantum graphs}
\label{sec:sqg}

The interest in quantum graphs started with the studies of Kottos and
Smilansky \cite{PRL.79.4794.1997,AoP.274.76.1999} in the context of
quantum chaos, where they analyzed the spectral statistics of simple
quantum graphs and showed that the spectra closely follows the
prediction of the random matrix theory.
An important result from the studies of quantum graphs is the
possibility of obtaining analytical solutions even when they present
chaotic behavior
\cite{PRL.88.044101.2002,PRE.65.046222.2002,PE.9.523.2001,
PRE.64.036225.2001}.
The spectral analysis in graphs is still of current interest
\cite{CSF.156.111817.2022}, and quantum graphs are deeply related to the
concept of quantum walks \cite{CP.44.307.2003} as discussed by Tanner in
Ref. \cite{Incollection.2006.Tanner}.
Since quantum walks are the quantum version of the classical random
walks, the case of classical random walks is also of interest.
As one knows, classical random walks are diffusive, so diffusion in
graphs and networks is also of current interest
\cite{CSF.156.111791.2022} and may be investigated using quantifiers of
entropy as well.

In order to develop our investigation, we recall that, mathematically,
a simple quantum graph, $\Gamma(V,E)$, consists of
(i) a metric graph, i.e., a set of $v$ vertices, $V=\{1,\ldots,v\}$, and
a set of $e$ edges, $E=\{e_1,\ldots,e_e\}$, where each edge is a pair of
vertices $e_s=\{i,j\}$, and we assign positive lengths to each edge
$\ell_{e_{s}}\in (0,\infty)$.;
(ii) a differential operator, $H$; and
(iii) a set of boundary conditions (BC) at the vertices $i$ of the graph
\cite{Book.Berkolaiko.2012}.
Thus we say that a quantum graph is a triple
$\{\Gamma(V,E), H,\text{BC}\}$.
In this work, we consider the free Schr\"odinger operator
$H=-(\hbar^2/2m)d^2/dx^2$ on each edge and, if not explicitly stated,
Neumann boundary conditions (also called standard vertex conditions) on
the vertices.
The graph topology is totally defined by its adjacency matrix
$A(\Gamma)$ of dimension $v \times v$, whose elements are given by
$A_{ij}=1$ if $i$ is adjacent to $j$ and $0$ otherwise.
We create an open quantum graph, $\Gamma^{l}$, which is suitable to
study scattering problems, by adding $l$ leads (semi-infinite edges) to
its  vertices (see Fig. \ref{fig:fig1}).
The open quantum graph $\Gamma^{l}$ then represents a scattering system
with $l$ scattering channels which is characterized by the energy
dependent global scattering matrix
$\boldsymbol{\sigma}_{\Gamma^{l}}(k)$, where $k=\sqrt{2mE/\hbar^2}$ is
the wave number and the matrix elements are given by the scattering
amplitudes $\sigma_{\Gamma^{l}}^{(f,i)}(k)$, where $i$ and $f$ are the
entrance and exit scattering channels, respectively.

In this work, we employ the Green's function approach as developed in
Ref. \cite{PRA.98.062107.2018} to determine the scattering amplitudes $\sigma_{\Gamma^{l}}^{(f,i)}(k)$.
This technique was used to study narrow peaks of full transmission and
transport in simple quantum graphs
\cite{PRA.100.62117.2019,EPJP.135.451.2020}, and the scattering Shannon
entropy for quantum graphs \cite{PE.141.115217.2022,PRA.103.062208.2021}.
These latter ones have inspired us to describe the present study.
The exact scattering Green's function for a quantum particle of wave
number $k$ entering the graph at the vertex $i$ and exiting at the
vertex $f$ is given by
\begin{equation}
  \label{eq:GF}
  G_{\Gamma^{l}}^{(f,i)}(k) = \frac{m}{i \hbar^2 k}
  \left[
  \delta_{fi} e^{i k |x_f-x_i|}+
  \sigma_{\Gamma^{l}}^{(f,i)}(k) e^{i k (|x_{f}|+|x_{i}|)}
  \right],
\end{equation}
where
\begin{equation}
  \label{eq:Srs}
  \sigma_{\Gamma^{l}}^{(f,i)}(k) =
  \delta_{fi}r_{i}
  +
  \sum_{j \in E_{i}}  A_{ij} P_{ij}^{(f)}t_{i},
\end{equation}
which are the global scattering amplitudes.
In Eq. \eqref{eq:GF}, $x_i$ and $x_f$ are reference points on the leads
connected to the vertices $i$ and $f$, respectively.
In Eq. \eqref{eq:Srs}, $E_{i}$ is the set of neighbor vertices
connected to $i$, and  $r_{i}$ ($t_{i}$) is the $k$-dependent
reflection (transmission) amplitude at the vertex $i$, which is
determined unambiguously from the BC.
The so-called family of paths $P_{ij}^{(f)}$ between the vertices $i$
and $j$, are given by
\begin{equation}
  \label{eq:pij}
  P_{ij}^{(f)}
  =
   z_{ij}\delta_{fj} t_{j}
  +
  z_{ij}  P_{ji}^{(f)} r_{j}
  +z_{ij} \sum_{l \in {E_{j}^{i,f}}}  A_{jl} P_{jl}^{(f)} t_{j},
\end{equation}
where $z_{ij}= e^{i k \ell_{s}}$ with $\ell_{s}$ being the length of the
edge $e_{s}=\{i,j\}$ connecting $i$ and $j$, and $E_{j}^{i,f}$ being the
set of neighbors vertices of $j$ but with the vertices $i$ and $f$
excluded.
The family $P_{ji}^{(f)}$ can be obtained from the above equation by
swapping $i \leftrightarrow j$, and the number of family of paths is
always twice the number of edges in the underlying graph.
For a given quantum graph, solving the inhomogeneous system of equations
provided by Eq. \eqref{eq:pij} we obtain the final form for the exact
Green's function, from which we can easily extract the global scattering
amplitudes for the underlying quantum graph.
It is important to stress that the above procedure is based on the
solution of the system of equations provided by Eq. \eqref{eq:pij},
which is obtained from the adjacency matrix of the graph, and replaces
the cumbersome sum over paths of Ref. \cite{PR.647.1.2016}, being
mathematically equivalent to the approach used in
\cite{JPA.56.475202.2023}.

In general, the specific boundary condition imposed at the vertex $i$
defines the individual scattering amplitudes, $r_i$ and $t_i$, in such a
way that they are $k$-dependent (energy-dependent).
However, for a vertex $i$ of degree $d_i \geq 2$ with Neumann boundary
condition, the scattering amplitudes are given by
\begin{equation}
  \label{eq:scatt_amp}
  r_{i} = \frac{2}{d_{i}} -1,
  \quad \text{and} \quad
  t_{i} = \frac{2}{d_i},
\end{equation}
which are $k$-independent and the vertex is usually called a Neumann
vertex.
It is also important to observe that even though the above individual
quantum amplitudes are $k$-independent the global scattering amplitudes
$\sigma_{\Gamma^{l}}^{(f,i)}(k)$ are $k$-dependent.
For the case where the vertex is a dead end having degree $1$, the
Neumann and  Dirichlet boundary conditions lead to $r_i=1$ and $r_i=-1$,
respectively \cite{PR.647.1.2016}.

Consider now again the scattering quantum graph $\Gamma^{l}$ described
in the previous section.
By fixing the entrance channel, say $i$, this scattering system is
characterized by $l$ quantum amplitudes, which are obtained
\textit{analytically} from the Green's function in Eq. \eqref{eq:GF},
and defines a set of $l$ quantum probabilities as
\begin{equation}
  p_{\Gamma^{l}}^{(j,i)}(k)=|\sigma_{\Gamma^{l}}^{(j,i)}(k)|^{2}.
\end{equation}
These are probabilities for a particle entering the graph, with a wave
number $k$, by the fixed vertex $i$ and exiting the graph by the vertex
$j$ (including the vertex $i$ itself), and fulfills the relation
\begin{equation}
  \sum_{j=1}^{l}p_{\Gamma^{l}}^{(j,i)}(k)=1,
\end{equation}
to ensure unitarity.
Then, when a scattering process occurs in a quantum graph, it provides
$l$ distinct probabilities (the scattering probabilities), in a way
similar to a discrete random variable with $l$ possible outcomes.
Thus, in \cite{PRA.103.062208.2021,PE.141.115217.2022} based on the
Shannon entropy, we introduced the scattering Shannon entropy for
quantum graphs, which is given by
\begin{equation}
  H_{\Gamma^l}^{(i)}(k)=
  -\sum_{j=1}^{l} p_{\Gamma^{l}}^{(j,i)}(k)
  \log_2 \left[ p_{\Gamma^{l}}^{(j,i)}(k)\right].
\end{equation}
The above quantity encodes the informational content as a function of
$k$ of the scattering process in a graph.
Moreover, for graphs which have periodic scattering probabilities with
period $K$, we can compute the average scattering entropy, which is
written as
\begin{equation}
  \label{eq:AS}
  \bar{H}_{\Gamma^{l}}^{(i)}=\frac{1}{K}
  \int_0^K H_{\Gamma^{l}}^{(i)} (k)\, dk,
\end{equation}
which encodes all the complicated $k$-dependent behavior of the quantum
probabilities along the period $K$ to a global information which is
independent of $k$.

\begin{figure}[t]
  \centering
  \includegraphics[width=\threedpwidth]{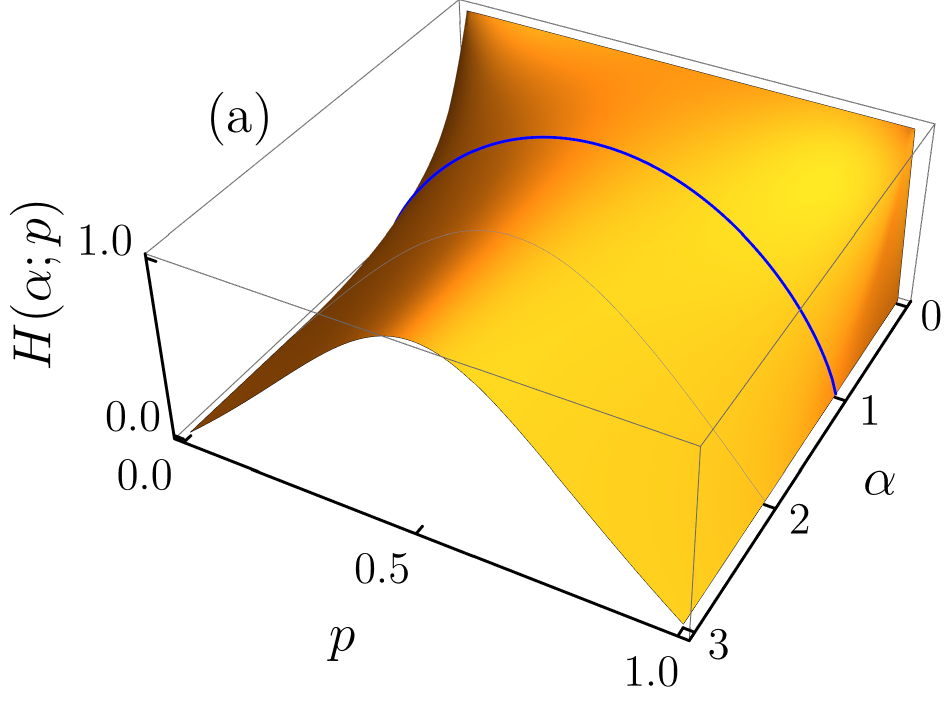}
  \quad
  \includegraphics[width=\threedpwidth]{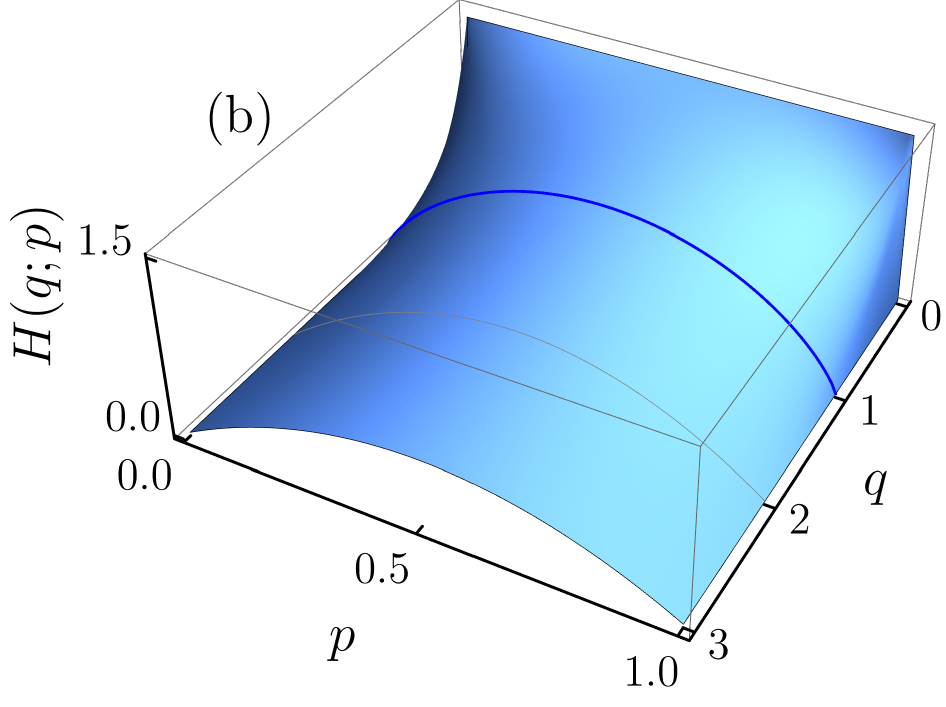}
  \caption{
     (a) R\'enyi entropy for a Bernoulli random variable with
     probabilities $p$ and $(1-p)$ as a function of
     $p$ and $\alpha$
     and
     (b) Tsallis entropy for a Bernoulli random variable with
     probabilities $p$ and $(1-p)$ as function of $p$ and $q$.
     The dark blue curves show the result of the Shannon entropy.}
  \label{fig:fig2}
\end{figure}

In the present work, we generalize this concept to both the R\'enyi and
Tsallis entropies. In this manner, the
\textit{scattering R\'enyi entropy}, which is parametrized by $\alpha
\geq 0$ and $\alpha\neq 1$, is defined as
\begin{equation}
  H_{\Gamma^{l}}^{(i)}(\alpha;k)=
  \frac{1}{1-\alpha} \log_2 \sum_{j=1}^l
  \left[p_{\Gamma^l}^{\left(j,i\right)}(k)\right]^{\alpha}.
\end{equation}
In the limit $\alpha \to 1$, this result converges to the scattering
Shannon entropy. In a similar way, the \textit{scattering Tsallis
  entropy}, which is  parametrized by $q\geq 0$ and $q\neq 1$, is
defined as
\begin{equation}
   H_{\Gamma^{l}}^{(i)}(q;k)=
  \frac{\log_2 e}{q-1}
  \left\{
    1-\sum_{j=1}^l \left[p_{\Gamma^l}^{\left(j,i\right)}(k)\right]^{q}
  \right\},
\end{equation}
which also converges to the scattering Shannon entropy for $q \to 1$.
Moreover, for graphs which have periodic scattering probabilities with
period $K$, we can compute the
\textit{average scattering R\'enyi entropy} and the
\textit{average scattering Tsallis entropy}, which are given by
\begin{equation}
  \label{eq:RT-AS}
  \bar{H}_{\Gamma^{l}}^{(i)}(\alpha)=\frac{1}{K}
  \int_0^K H_{\Gamma^{l}}^{(i)}(\alpha;k)\, dk,
\end{equation}
and
\begin{equation}
   \bar{H}_{\Gamma^{l}}^{(i)}(q)=\frac{1}{K}
  \int_0^K H_{\Gamma^{l}}^{(i)}(q;k)\, dk,
\end{equation}
respectively.

\begin{figure}[b]
  \centering
  \includegraphics[width=\threedpwidth]{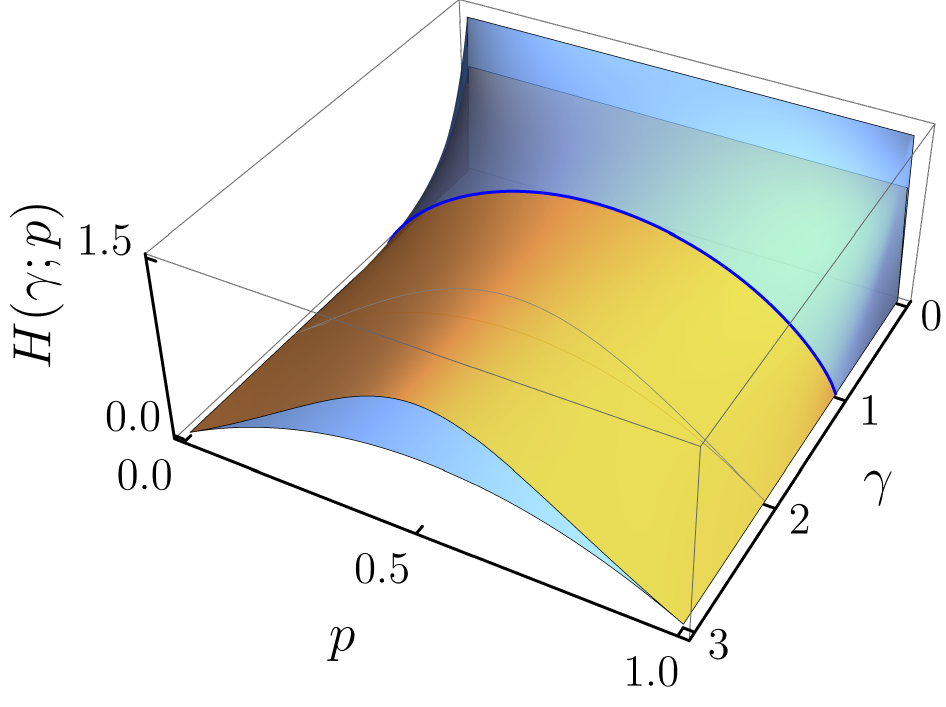}
  \caption{
    The R\'enyi (for $\gamma=\alpha$) and Tsallis (for $\gamma=q$)
    entropies that appear in Fig. \ref{fig:fig2}, displayed together to
    visually illustrate their quantitative differences.}
  \label{fig:fig3}
\end{figure}

In Fig. \ref{fig:fig2}(a) we show the behavior of the R\'enyi entropy
and in Fig.  \ref{fig:fig2}(b) the Tsallis entropy for a Bernoulli
random variable which has two possible outcomes, one with probability
$p$ and the other with probability $1-p$, as a function of $p$ and
$\alpha$ and $p$ and $q$, respectively.
We see that the two entropies are maximized at $p=1/2$, that is, in the
case with the two random variables having equal probabilities.
Also, as  $\alpha$ and $q$ approach unity, the two results converge to
the Shannon entropy, but when they are different, the two entropies
become more and more distinct.
In particular, the R\'enyi entropy is greater or equal to the Tsallis
entropy, when $\alpha=q \geq1$, but it is lower or equal to for
$\alpha=q\leq1$.
We illustrate this in Fig. \ref{fig:fig3}, where we display the results
of Fig. \ref{fig:fig2} in the same diagram, visually comparing the two
entropies explicitly.

We think it is of current interest to examine different quantum graphs,
with several arrangements of vertices and edges, to see how the results
appear in these new situations.
We do this in the next section, but we know that the R\'enyi and the
Tsallis entropies work to highlight different aspects of physical
systems, so in the present work, we do not want to compare these two
distinct quantifiers against each other.
Instead, we concentrate mainly on how the scattering R\'enyi entropy and
the average scattering R\'enyi entropy behave as one varies the
parameter $\alpha$, and how the scattering Tsallis entropy and the
average scattering Tsallis entropy change as one modifies the parameter
$q$.

\section{Examples}
\label{sec:examples}

Let us now calculate the scattering entropies for different types of
graphs \cite{Book.2010.Diestel}, when we add only two leads to the
graph, as we illustrate in Fig. \ref{fig:fig1}(b), and when we add
one lead to each vertex of the graph, as we illustrate in
Fig. \ref{fig:fig1}(c).
Interesting possibilities refer to the presence of several scattering
channels, where the scattering entropies provide a useful
simplification in the analysis, condensing the several possible results
in just  one.
There are some limiting cases that can be observed: one is when all
scattering channels are equivalent, which corresponds to the case where
the scattering has the highest informational value, resulting in the
maximum value for the scattering entropies; another is when the
scattering has one preferential channel, such that the one scattering
coefficient equals unity, leading to vanishing scattering entropies.
Below we illustrate several distinct possibilities.

To exemplify the main results of the previous section, let us now focus
on some specific cases of quantum graphs.
We start with some arrangements of graphs, like the series and the
parallel layouts.
Then, we examine results for specific types of graphs with $n$ vertices
and leads, or channels, connected to them.
In the last case, the types of graphs which we will consider are the
cycle ($C_n$), wheel ($W_n$) and the complete ($K_n$) on $n$ vertices.
In general, we calculate the scattering coefficients of these quantum
graphs using the Neumann boundary condition.

\subsection{Quantum graphs in series and parallel layouts}

\begin{figure}[t]
  \centering
  \includegraphics[width=1.2\gwidth]{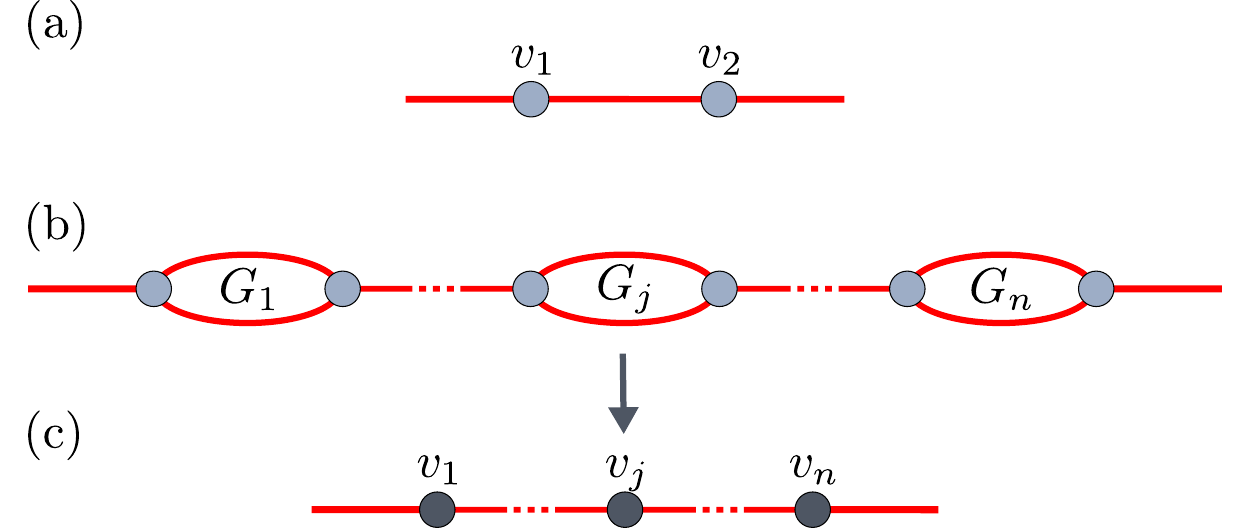}
  \caption{
    (a) A graph on $2$ vertices connected by an edge with one lead
    attached to each vertex, (b) a series arrangement of quantum graphs,
    $G_j$, consisting of two vertices connected by two edges, and (c) a
    series arrangement of vertices \{$v_1,\dots,v_j,\dots,v_n$\},
    each one is a renormalization of a respective quantum graph, $G_j$.
  }
  \label{fig:fig4}
\end{figure}

To introduce different arrangements of graphs, we can start with one of
the simplest graphs, which is defined by two vertices connected by one
edge (a path graph on $2$ vertices, $P_2$).
By adding one lead to each vertex, as shown in Fig. \ref{fig:fig4}(a),
and defining boundary conditions on them, we obtain the
reflection \{$r_1$,$r_2$\} and transmission \{$t_1$,$t_2$\} amplitudes
associated to the respectively vertices $v_1$ and $v_2$.
Furthermore, the metric in this quantum graph is defined by the length
$\ell$ of the edge which connects these vertices.

For notational ease in this subsection, the reflection and transmission
amplitudes will be denoted by $R$ and $T$ instead of $\sigma$.
Thus, defining $z=e^{i k \ell}$ the scattering amplitudes of this
quantum graph are \cite{PR.647.1.2016}
\begin{equation}
    R_{S(v_{1},v_{2})}=r_{1}+\frac{t_{1}^{2}r_{2}z^{2}}{1-r_{1}r_{2}z^{2}},
\end{equation}
and
\begin{equation}
    T_{S(v_{1},v_{2})}=\frac{t_{1}t_{2}z}{1-r_{1}r_{2}z^{2}},
\end{equation}
where $S(v_1,v_2)$ is used to identify that the vertices $v_1$ and $v_2$
are arranged in a series layout.

These equations can be used recursively to obtain the scattering
amplitudes when adding one more vertex ($v_3$) in series with the first
quantum graph, by  defining
$\{r_1,t_1\} \xrightarrow{} \{R_{S(v_1,v_2)},T_{S(v_1,v_2)}\}$ and
$\{r_2,t_2\} \xrightarrow{} \{r_3,t_3\}$.
In this sense, it is possible to obtain the scattering amplitudes for
the general case of $n \ge 3$ by transforming these equations to a
recursive equation system in the form
\begin{align}
\label{eq:Rseries}
  R_{S(v_{1},\dots,v_{n})}= {}
  &
    R_{S(v_{1},\dots,v_{n-1})}
    +\frac{T_{S(v_{1},\dots,v_{n-1})}^{2}r_{n}z^{2}}
    {1-R_{S(v_{1},\dots,v_{n-1})}r_{n}z^{2}},\\
    \label{eq:Tseries}
  T_{S(v_{1},\dots,v_{n})} = {}
  &
    \frac{T_{S(v_{1},\dots,v_{n-1})}t_{n}z}
    {1-R_{S(v_{1},\dots,v_{n-1})}r_{n}z^{2}}.
\end{align}
Here we have a set of $n$ vertices in a series layout,
$S(v_1,\dots,v_n)$, which allow us to study a set of $n$ renormalized
quantum graphs displayed in Fig. \ref{fig:fig4}(c).
Thus, with Eqs. \eqref{eq:Rseries} and \eqref{eq:Tseries}, we have an
alternative recursive method to the transfer matrix method for quantum
graphs \cite{JPA.33.3567.2000} arranged in this layout, which only needs
the scattering amplitudes of each quantum graph.

Now we can find the scattering amplitudes for the special case where $n$
quantum graphs $G$ with the same scattering amplitudes of reflection
($\mathcal{R}$) and transmission ($\mathcal{T}$) are arranged in a
series layout.
Thus, the final form of its reflection and transmission amplitudes are
\begin{equation}
    \label{eq:series_reflection}
    R_{S(v_1,\dots,v_{n})}=
    \frac{2\left(\Lambda_{+}^{n}-\Lambda_{-}^{n}\right)\mathcal{R}}
    {\left(2-\Lambda_{-}\right)\Lambda_{+}^{n}
      -\left(2-\Lambda_{+}\right)\Lambda_{-}^{n}},
\end{equation}
and
\begin{equation}
    \label{eq:series_transmission}
    T_{S(v_{1},\dots,v_{n})}=
    \frac{2^{n}\left(\Lambda_{+}-\Lambda_{-}\right)\mathcal{T}^{n}z^{n-1}}
    {\left(2-\Lambda_{-}\right)\Lambda_{+}^{n}
      -\left(2-\Lambda_{+}\right)\Lambda_{-}^{n}},
\end{equation}
with
\begin{align}
  \Lambda_{\pm} = {}
  &
  1+\left(\mathcal{T}^{2}-\mathcal{R}^{2}\right)z^{2}
                       \pm
  \left[
    1-2\left(\mathcal{R}^{2}+\mathcal{T}^{2}\right)z^{2}
    +\left(\mathcal{R}^{2}-\mathcal{T}^{2}\right)^{2}z^{4}
  \right]^{\frac{1}{2}}.
\end{align}

\begin{figure}[t]
  \centering
  \includegraphics[width=\gwidth]{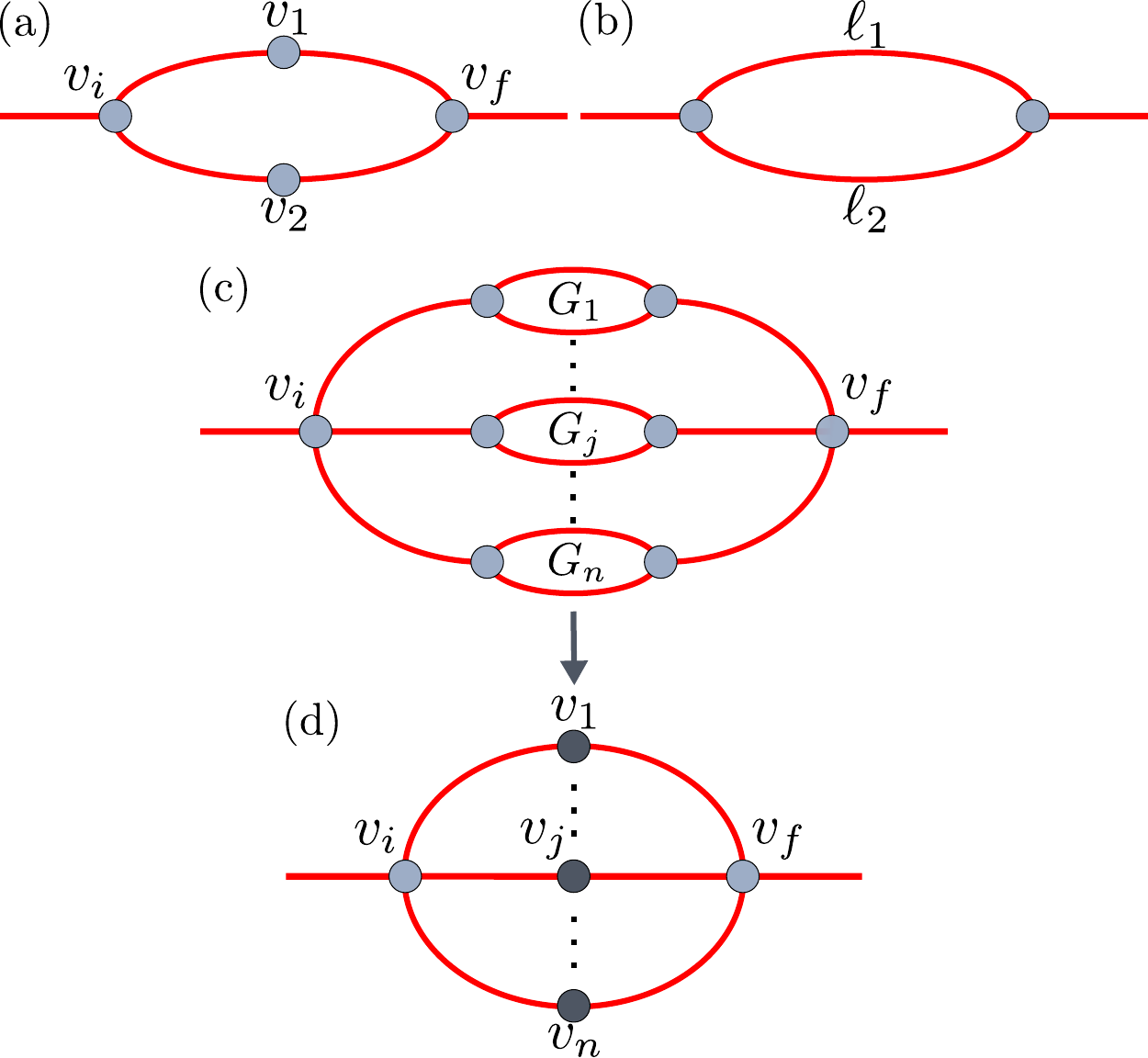}
  \caption{
    (a) Parallel arrangement of two vertices $v_1$ and $v_2$;
    (b) two vertices connected by two edges of length $\ell_1$ and
    $\ell_2$;
    (c) a set of $n$ $P(v,v)$ graphs in a parallel layout; and
    (d) a set of $n$ vertices, which are renormalizations (darker grey)
    of two edge graphs with the same length $P(v,v)$, in a parallel
    layout.
  }
  \label{fig:fig5}
\end{figure}

A second layout can be defined starting with two vertices $v_1$ and
$v_2$ in parallel as shown in Fig. \ref{fig:fig5}(a), where both
vertices are only connected to the other two vertices $v_i$ and $v_f$,
which are the vertices with leads attached to them.
Here we define the vertices $v_i$ and $v_f$ with the same amplitudes
$r$ and $t$, and $v_1$ and $v_2$ with the respectively scattering
amplitudes $r_1$, $t_1$ and $r_2$, $t_2$.
Furthermore, to simplify these equations, here we define
$z_j = e^{i k\ell_j}$, where $\ell_j$ is the distance between the
vertices in parallel to the ones with leads attached, as shown in
Fig. \ref{fig:fig5}(a).
So, we have the scattering amplitudes for this system
as
\begin{align}
  R_{P(v_{1},v_{2})}= {}
  &
    r
    + \frac{t^{2}}{\Lambda_{P(v_1,v_2)}}
    \Big\{
    C_{1,2}+C_{2,1}  +
    2\left[t t_{1} t_{2}-\left(2r-t\right)r_{1}r_{2}\right]z_{1}^{2}z_{2}^{2}
   \nonumber \\
 &
    -2\left(r+t\right)\left(r-t\right)^{2}
    \left(r_{1}^{2}-t_{1}^{2}\right)
    \left(r_{2}^{2}-t_{2}^{2}\right)z_{1}^{4}z_{2}^{4}
    \Big\},
    \label{eq:parallel_2_reflection}
  \\
  T_{P(v_{1},v_{2})}= {}
  &
    \frac{t^2}{\Lambda_{P(v_1,v_2)}}
    \Big\{t_{1}z_{1}^{2}+t_{2}z_{2}^{2}
    -2\left(r-t\right)\left(r_{1}t_{2}+r_{2}t_{1}\right)z_{1}^{2}z_{2}^{2}
    \nonumber \\
 &
    +\left(r-t\right)^{2}
    \big[
    \left(r_{1}^{2}-t_{1}^{2}\right)t_{2}z_{1}^{2}
    +t_{1}\left(r_{2}^{2}-t_{2}^{2}\right)z_{2}^{2}
    \big]
    z_{1}^{2}z_{2}^{2}\Big\},
    \label{eq:parallel_2_transmission}
\end{align}
where
\begin{align}
  \Lambda_{P(v_1,v_2)} = {}
  &
    \Big[1-r\left(r_{1}z_{1}^{2}+r_{2}z_{2}^{2}\right)
    +\left(r^{2}
    -t^{2}\right)\left(r_{1}r_{2}+t_{1}t_{2}\right)z_{1}^{2}z_{2}^{2}\Big]^{2}
 \nonumber \\
 &
    -\Big[
    r\left(t_{1}z_{1}^{2}+t_{2}z_{2}^{2}\right)
    -\left(r^{2}-t^{2}\right)
    \left(r_{1}t_{2}+t_{1}r_{2}\right)z_{1}^{2}z_{2}^{2}
    \Big]^{2},
\end{align}
and
\begin{align}
  C_{m,n} = {}
  &
    r_{m}z_{m}^{2}-r\left(r_{m}^{2}-t_{m}^{2}\right)z_{m}^{4}
    +r\left(r-t\right)r_{m}
    \left(r_{n}^{2}-t_{n}^{2}\right)z_{m}^{2}z_{n}^{4}
    +\left(2r^{2}-rt-t^{2}\right)
    \left(r_{m}^{2}-t_{m}^{2}\right)r_{n}z_{m}^{4}z_{n}^{2}.
\end{align}
To identify this parallel layout, we now use $P(v_1,v_2)$.

With these equations, we can approach the case where all the vertices
have a Neumann boundary condition
($r=-{1}/{3}$, $t={2}/{3}$, $r_1=r_2=0$, $t_1=t_2=1$).
Letting {$z_1 = e^{i k \ell_1/2}$ and $z_2= e^{i k \ell_2/2}$}, we obtain
the equations for the quantum graph in Fig. \ref{fig:fig5}(a), which has
scattering amplitudes
\begin{equation} R_{P(v,v)}=
  -\frac
  {3+\left(z_1^2-z_2^2\right)^2-3\left(2-z_1^2z_2^2\right)z_1^2z_2^2}
  {
    \left(3-z_1^2z_2^2\right)^{2}
    -\left(z_1^2+z_2^2\right)^{2}},
\end{equation}
and
\begin{equation} T_{P(v,v)}=
  \frac
  {4
    \left[
      \left(1-z_2^4\right)z_1^2
      +\left(1-z_1^4\right)z_2^2
    \right]
  }
  {
   \left(3-z_1^2z_2^2\right)^{2}
    -\left(z_1^2+z_2^2\right)^{2}
  }.
\end{equation}
These equations allow us to study quantum interference due to the
difference in the length of its edges.
Works that have used these graphs have proposed the use of them in
a series layout as quantum devices \cite{EPJP.135.451.2020,EMCE.52599.2021}.
Here we use the particular case where both edges have the same length
($\ell_1=\ell_2=\ell$ and $z_1^2=z_2^2=z$), which leads to the quantum
amplitudes
\begin{equation}
    \label{eq:parallel_samelengths_reflection}
    \mathcal{R} = -\frac{3-3 z^{2}}{9-z^{2}},
\end{equation}
\begin{equation}
    \label{eq:parallel_samelengths_transmission}
    \mathcal{T} = \frac{8z}{9-z^{2}}.
\end{equation}

We can generalize the model of quantum graphs in a parallel layout by
defining a set of $n$ quantum graphs that are connected only to the same
two vertices, as illustrated in the Fig. \ref{fig:fig5}(c).
As these quantum graphs can be renormalized to vertices, each one having
the same scattering amplitudes of the original quantum graph $G_j$.
Here we extend the Eqs. \eqref{eq:parallel_2_reflection} and
\eqref{eq:parallel_2_transmission} to the case where we have $n$
quantum graphs $G_j$ in this parallel layout.
An interesting result considers the model where all the quantum graphs in
parallel have the same scattering amplitudes $\mathcal{R}$ and
$\mathcal{T}$, and the two lateral vertices have amplitudes $r$ and
$t$.
So, this model leads to scattering amplitudes in the form
\begin{align}
  R_{P(v_{1},...,v_{n})}= {}
  &
  r+\frac{nt^{2}}{\Lambda_{P(v_1,\ldots,v_n)}}
  \Big\{
  \mathcal{R}z^2
  -\left[r+\left(n-1\right)t\right]
  \left(\mathcal{R}^{2}-\mathcal{T}^{2}\right)z^{4}
  \Big\},
\end{align}
and
\begin{equation}
  T_{P(v_{1},...,v_{n})}=
  \frac{ nt^{2}\mathcal{T}z^{2}}
  {
    \Lambda_{P(v_1,\ldots,v_n)}
  }.
\end{equation}
where
\begin{align}
  \Lambda_{P(v_1,\ldots,v_n)} = {}
  &
    \left\{1-\left[r+\left(n-1\right)t\right]\mathcal{R}z^{2}\right\}^{2}
    -\left\{\left[r+\left(n-1\right)t\right]\mathcal{T}z^{2}\right\}^{2}.
\end{align}

\begin{figure}[t]
  \centering
  \includegraphics[width=\twodpwidth]{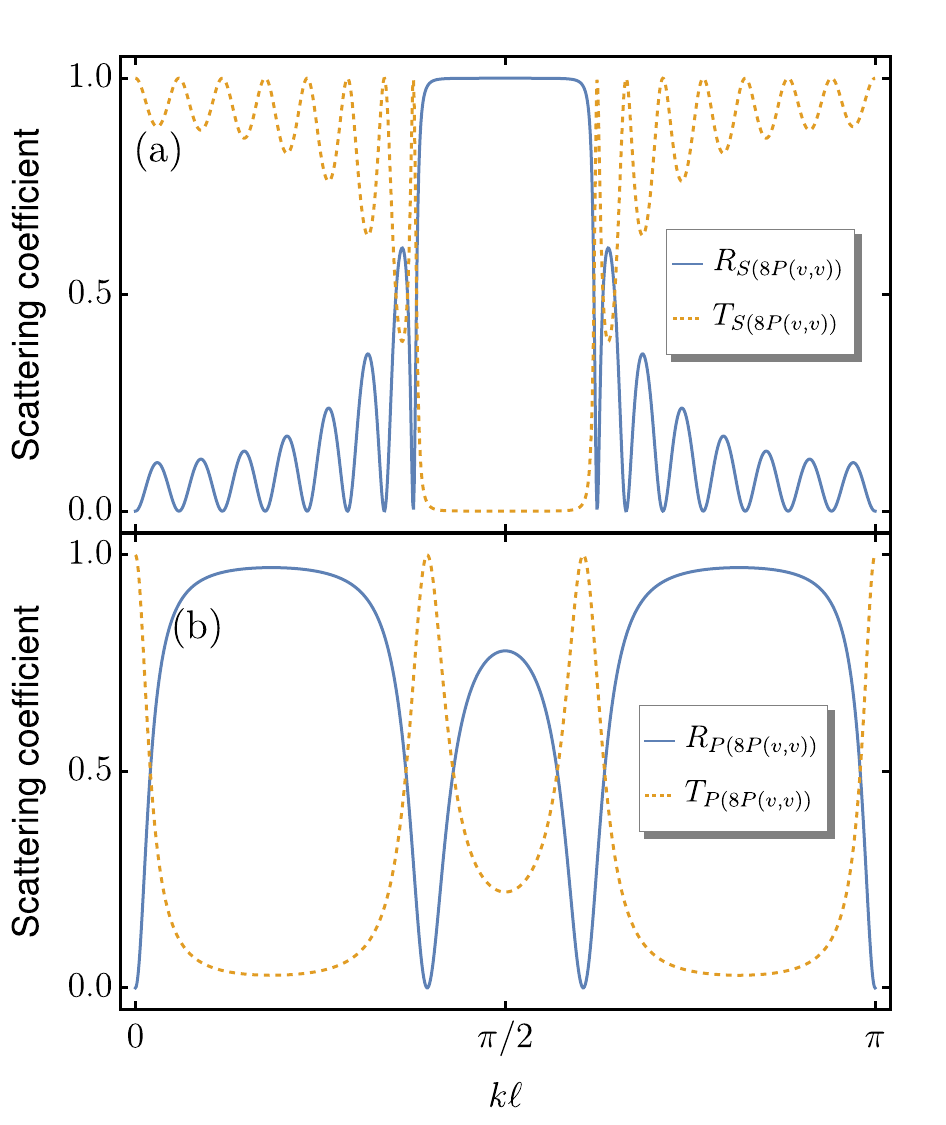}
  \caption{
   The reflection (blue) and transmission (yellow) coefficients of the
   (a) series and (b) parallel layout, with 8 quantum graphs
   $P(v,v)$.
  }
  \label{fig:fig6}
\end{figure}

\begin{figure}[t]
  \centering
  \includegraphics[width=\threedpwidth]{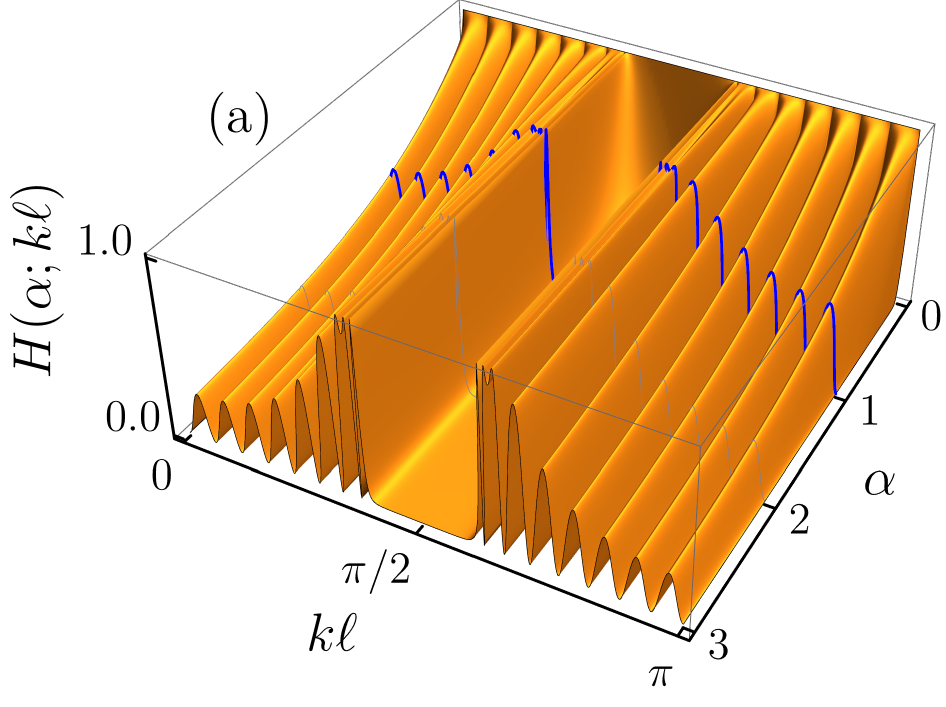}
  \quad
  \includegraphics[width=\threedpwidth]{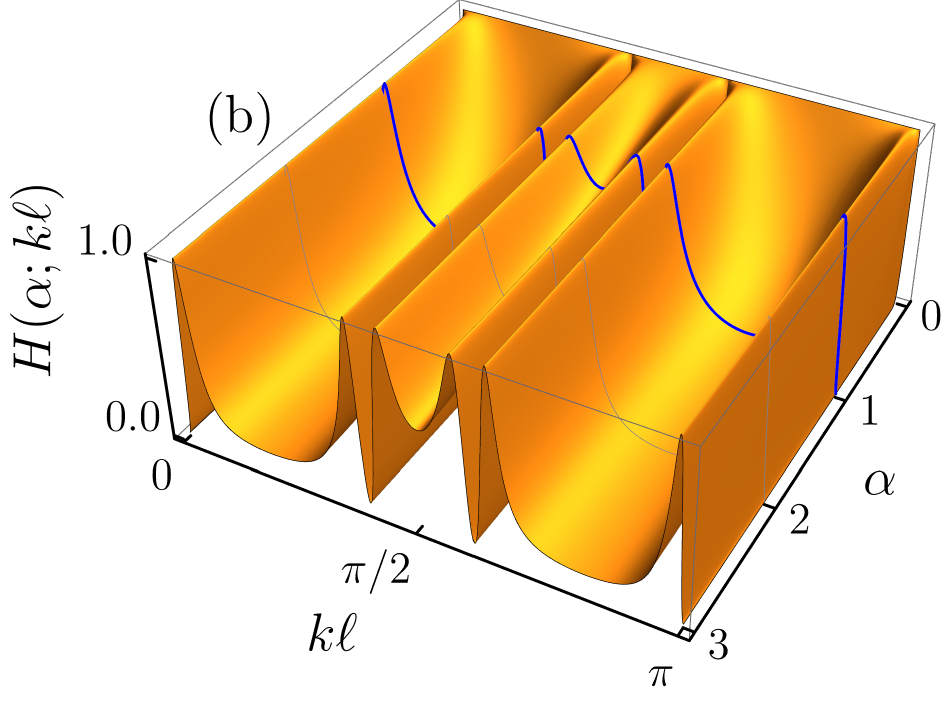}\\
  \includegraphics[width=\threedpwidth]{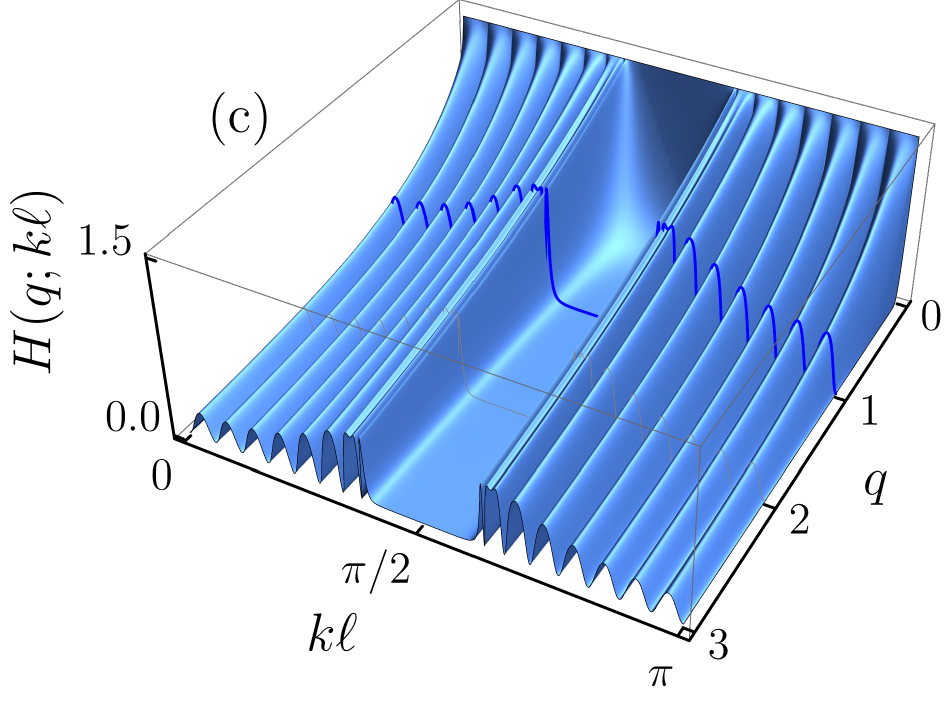}
  \quad
  \includegraphics[width=\threedpwidth]{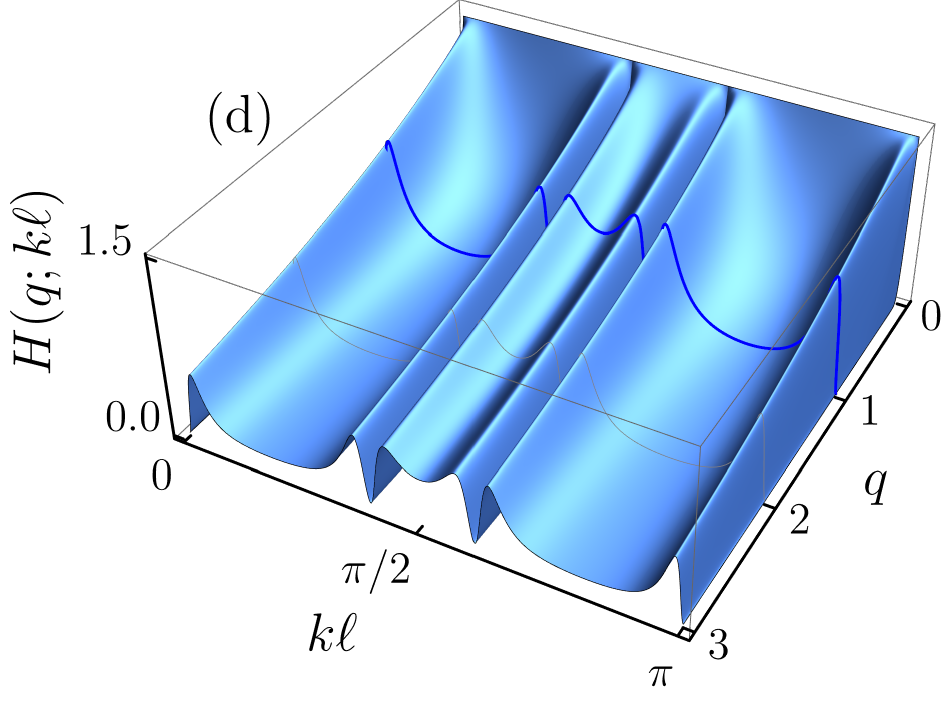}
  \caption{
    Scattering entropies for $8$ quantum graphs $P(v,v)$ in different
    layouts.
    R\'enyi scattering entropies for (a) series and (b) parallel layouts.
    Tsallis scattering entropies for (c) series and (d) parallel layouts.
    The limit for the Shannon entropy is highlighted in dark blue.
  }
  \label{fig:fig7}
\end{figure}

Finally, setting the Neumann boundary condition in the vertices
connected to the leads, here we have the amplitudes $r=-{(n-1)}/{(n+1)}$
and $t={2}/{(n+1)}$, then the final result is
\begin{align}
  \label{eq:parallel_reflection}
  \tilde{R}_{P(v_{1},...,v_{n})}= {}
  &
  \frac
  {
    2\left(n^{2}+1\right)\mathcal{R}z^{2}
    -\left(n^{2}-1\right)\left[1+\left(\mathcal{R}^{2}
        -\mathcal{T}^{2}\right)z^{4}\right]
  }
  {\tilde{\Lambda}_{P_{(v_1,\ldots,v_n})}},
\end{align}
and
\begin{equation}
  \label{eq:parallel_transmission}
  \tilde{T}_{P(v_{1},...,v_{n})}=
  \frac
  {4n\mathcal{T}z^{2}}
  {\tilde{\Lambda}_{P_{(v_1,\ldots,v_n})}},
\end{equation}
with
\begin{equation}
  \tilde{\Lambda}_{P_{(v_1,\ldots,v_n})}=
  \left[n+1-\left(n-1\right)\mathcal{R}z^{2}\right]^{2}
    -\left(n-1\right)^{2}\mathcal{T}^{2}z^{4}.
\end{equation}
To show how is the behavior of the scattering from the same set of
quantum graphs, but organized in different layouts, we considered a set
of eight of the same quantum graphs showed in Fig. \ref{fig:fig5}(b) with
$\ell_1=\ell_2=\ell$, which has the scattering amplitudes obtained in
Eqs. \eqref{eq:parallel_samelengths_reflection}
and \eqref{eq:parallel_samelengths_transmission}.
The series layout for this case is illustrated in Fig. \ref{fig:fig4}(b)
and has the scattering amplitudes given in
Eqs. \eqref{eq:series_reflection} and \eqref{eq:series_transmission},
being the reflection and transmission coefficients illustrated in
Fig. \ref{fig:fig6}(a).
For the parallel arrangement illustrated in Fig. \ref{fig:fig5}
(c), we obtained the coefficients from
Eqs. \eqref{eq:parallel_reflection} and
\eqref{eq:parallel_transmission}, which behaviors are illustrated in
Fig. \ref{fig:fig6}(b).
Finally, the R\'enyi (Tsallis) scattering entropies to the examples from
Fig. \ref{fig:fig6} are illustrated in Fig. \ref{fig:fig7}(a)
(Fig. \ref{fig:fig7}(c)) and
Fig. \ref{fig:fig7}(b) (Fig. \ref{fig:fig7}(d)), for both series and
parallel arrangements, respectively.

\subsection{Cycle quantum graph with $n$ leads}

\begin{figure}[b]
  \centering
  \includegraphics[width=\gwidth]{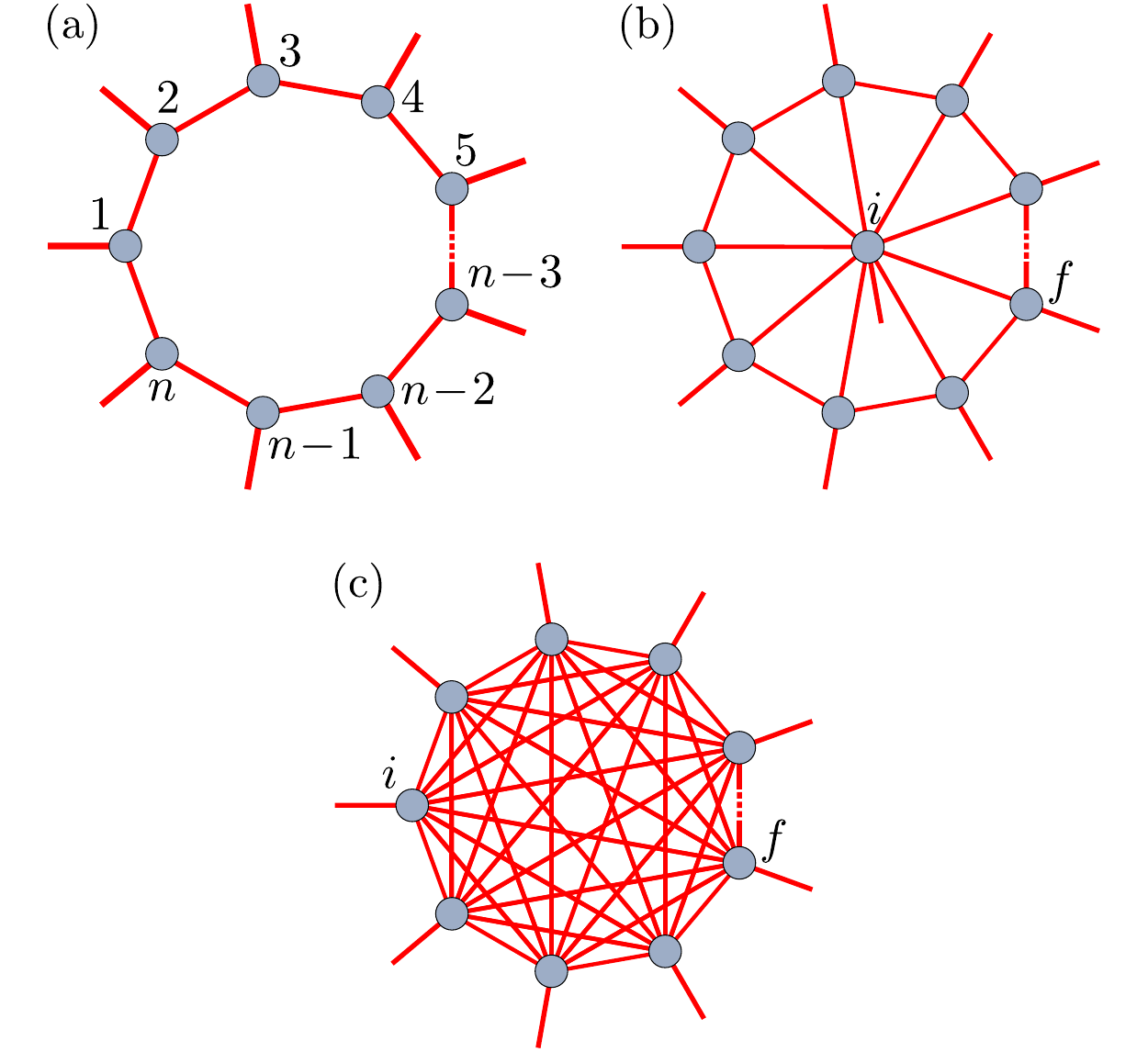}
  \caption{
    (a) Cycle, (b) wheel and (c) complete graphs  with leads attached to
    their vertices.
  }
  \label{fig:fig8}
\end{figure}

Cycle graphs, $C_n$, are defined as a set of $n$ vertices, each one with
degree two, therefore all the vertices have two neighbor vertices.
To define the scattering in the cycle quantum graph, we need to define
the entrance and the set of exit vertices, which have leads
attached to them, thus increasing their degrees by one.
In the present case, we define that every vertex in these graphs is
connected to one lead, being the vertex labeled as $1$ the entrance
vertex, and the remaining $n-1$ vertices, labeled from $2$ to $n$, are
the exit channels; see Fig. \ref{fig:fig8}.

To simplify the final form of equations in this subsection, we
define the quantities $\beta$ and $\mu_{\pm}$ in the following form
\begin{equation}
    \beta=\sqrt{9-10 z^{2}+z^{4}},
\end{equation}
and
\begin{equation}
    \mu_{\pm}=3+z^{2}\pm \beta.
\end{equation}
Thus, for a $C_n$ quantum graph with $n$ leads, we can write the
reflection amplitude as
\begin{align}
  \sigma_{C_{n}^{n}}^{\{1,1\}}= {}
  &
    \frac{
    \sqrt{\mu_{+}^{n}\mu_{-}}
    \left[\mu_{+}+\left(2+3\mu_{-}+4z-6z^{2}\right)z\right]
 -\sqrt{\mu_{+}\mu_{-}^{n}}\left[\mu_{-}+\left(2+3\mu_{+}+4z-6z^{2}\right)z\right]
    }
    {\Lambda_{\Gamma_{C_{n}}^{n}}^{\rm odd}},
\end{align}
with
\begin{equation}
  \Lambda_{C_{n}^{n}}^{\rm odd}=
  {\left(3-z\right)\left[\sqrt{\mu_{+}\mu_{-}^{n}}
    \left(\mu_{-}+4z\right)-\sqrt{\mu_{+}^{n}\mu_{-}}
    \left(\mu_{+}+4z\right)\right]},
  \end{equation}
for odd $n$, and
\begin{align}
  \sigma_{C_{n}^{n}}^{\{1,1\}}= {}
  &
    \frac
    {
    \mu_{+}\sqrt{\mu_{-}^{n}}\left[3\mu_{-}
    +\left(14+3\mu_{+}-6z^{2}\right)z^{2}\right]
    -\mu_{-}\sqrt{\mu_{+}^{n}}
    \left[3\mu_{+}+\left(14+3\mu_{-}-6z^{2}\right)z^{2}\right]
    }
     {
    \Lambda_{\Gamma_{C_{n}}^{n}}^{\rm even}
    },
\end{align}
with
\begin{equation}
  \Lambda_{C_{n}^{n}}^{\rm even}=
  {
    \left(9-z^{2}\right)\mu_{+}\mu_{-}\left(\sqrt{\mu_{+}^{n}}
    -\sqrt{\mu_{-}^{n}}\right)
    },
\end{equation}
for even $n$.
For the transmission amplitude, there is a dependence on which one
is the exit vertex, so we label the two first neighbors of the entrance
vertex ($1$) as the vertices $2$ and $n$, the two second neighbors as
the vertices $3$ and $n-1$, and so on, as shown in Fig. \ref{fig:fig8}(a).
Hence, the transmission amplitudes are defined in terms of the
chosen vertex $v$, with $2 \leq v \leq n$.
Thus, we have
\begin{align}
  \sigma_{C_{n}^{n}}^{\{v,1\}}= {}
  &
    \frac
    {2^{2v-3}z^{(v-2)}\left(1+z\right)\left(\mu_{+}\mu_{-}\right)^{-v}}
    {\Lambda_{C_{n}^{n}}^{\rm odd}}
    \Big[
    \left(4z-\mu_{+}\right)\sqrt{\mu_{+}^{2v+1}\mu_{-}^{n+4}}
    -\left(4z-\mu_{-}\right)\sqrt{\mu_{+}^{n+4}\mu_{-}^{2v+1}}
    \Big]
\end{align}
for odd $n$, and
\begin{align}
  \sigma_{C_{n}^{n}}^{\{v,1\}}= {}
  &
    \frac{2^{4v-1}z^{\left(v-1\right)}\beta\mu_{+}\mu_{-}
    \left(\sqrt{\mu_{+}^{n+2-2v}}+\sqrt{\mu_{-}^{n+2-2v}}\right)}
    {\Lambda_{C_{n}^{n}}^{\rm even}}.
\end{align}
for even $n$.

\begin{figure}[t]
  \centering
  \includegraphics[width=\twodpwidth]{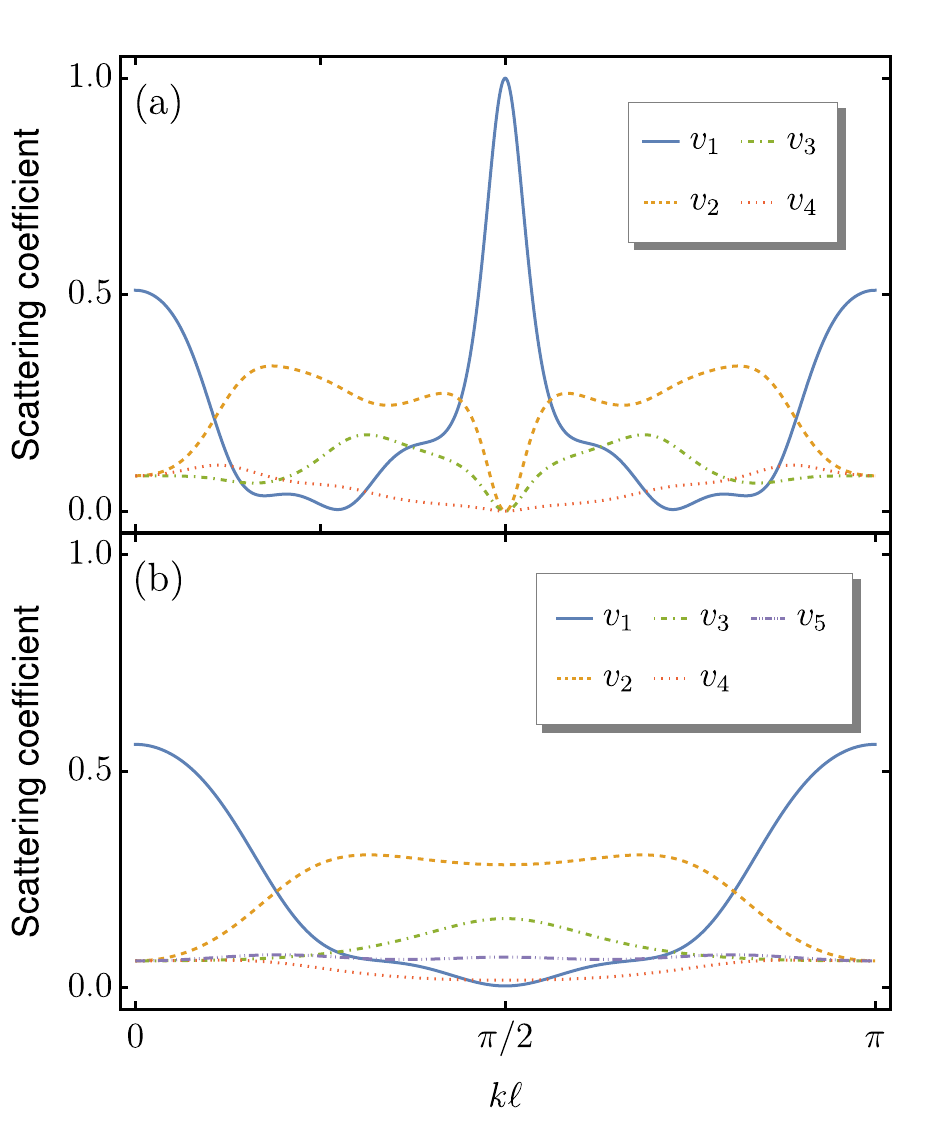}
  \caption{
   The reflection (blue) and transmission coefficients for cycle quantum
   graphs with (a) $7$ and (b) $8$ vertices.
  }
  \label{fig:fig9}
\end{figure}

As we considered all the vertices having the same boundary conditions,
it leads to a vertex $j$ be equivalent to another vertex $n-j+2$ with $2
\leq j \leq {(n+1)}/{2}$ for odd $n$, and $2 \leq j \leq {n}/{2}$ for
even $n$.
In this way, the number of scattering channels for an odd number of
vertices is reduced by ${(n-1)}/{2}$, and the reduction is of
${(n-2)}/{2}$ for even $n$.
So, for such graphs we have $n$ scattering amplitudes and due to the
symmetry of them, for the odd graphs there are ${(n+1)}/{2}$ different
values of them and for even $n$, the number of different scattering
amplitudes is ${(n+2)}/{2}$, as we exemplify in the Fig. \ref{fig:fig9} .

\begin{figure*}[t]
  \centering
  \includegraphics[width=\threedpwidth]{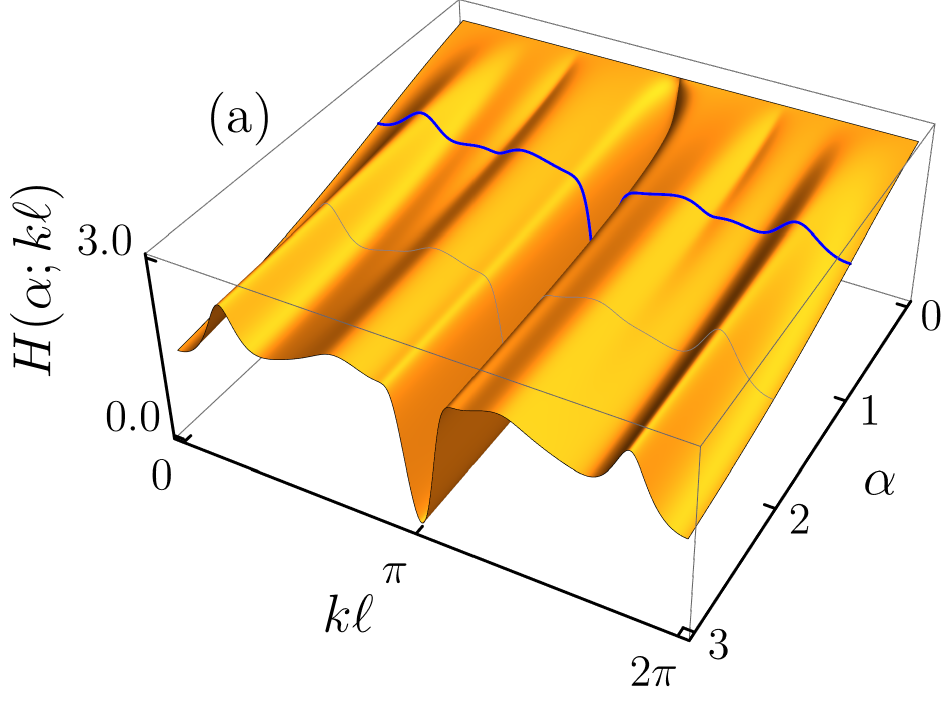}
  \quad
  \includegraphics[width=\threedpwidth]{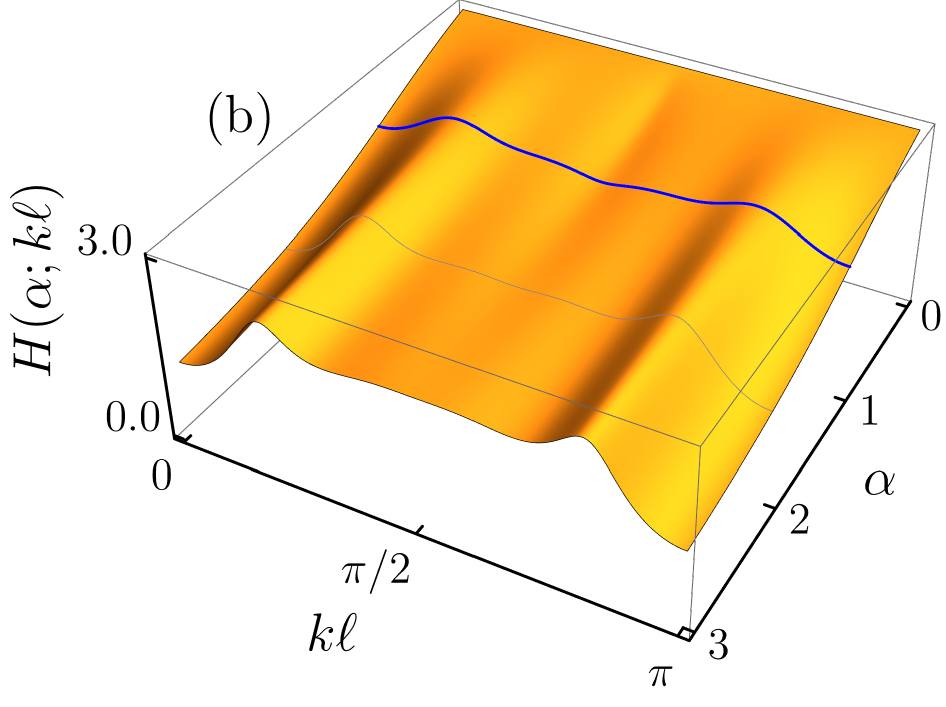}
  \\
  \includegraphics[width=\threedpwidth]{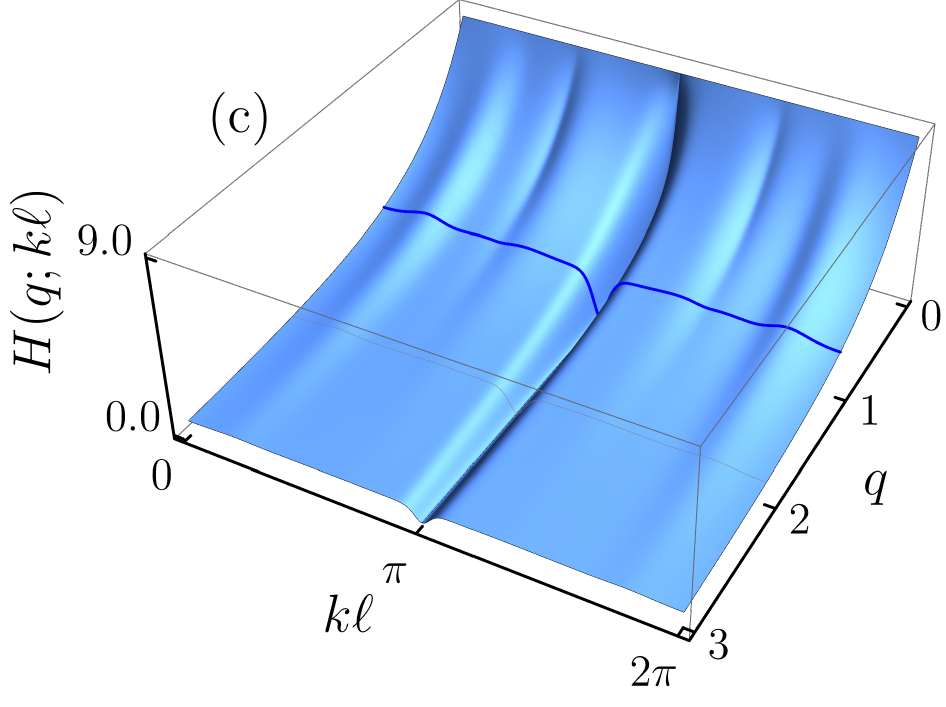}
  \quad
  \includegraphics[width=\threedpwidth]{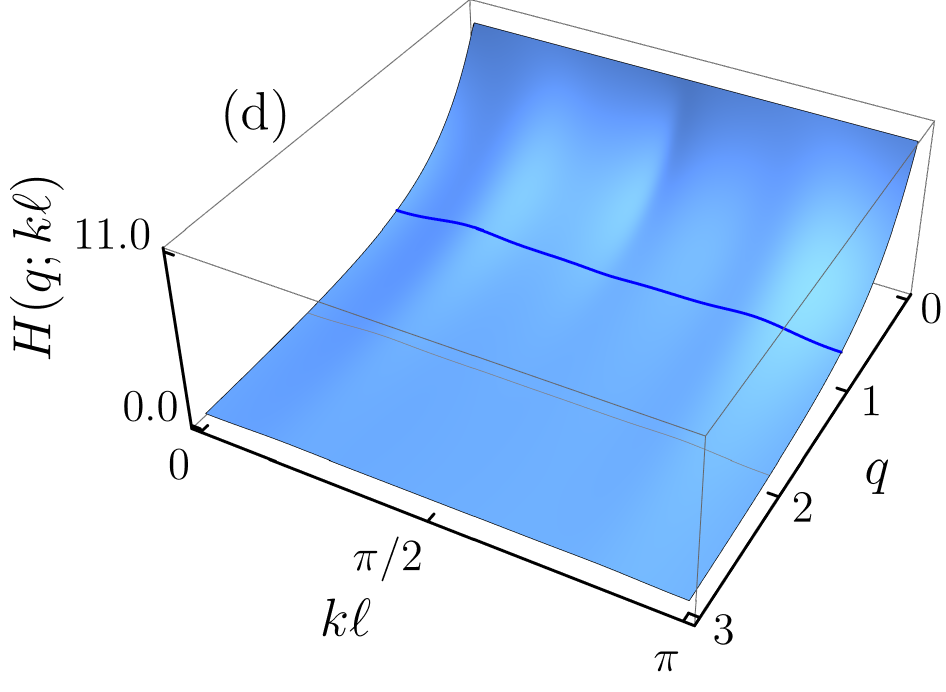}
  \caption{
    Scattering R\'enyi entropies for cycle quantum graph with (a) $7$
    and (b) $8$ vertices.
    Tsallis scattering entropies for cycle quantum graph with (c) $7$
    and (d) $8$ vertices.
   The limit for the Shannon entropy is highlighted in dark blue.
 }
  \label{fig:fig10}
\end{figure*}

In this sense, considering all the $n$ scattering coefficients that we
have obtained, we can obtain the scattering entropies which are
illustrated in Figs. \ref{fig:fig10}(a) and \ref{fig:fig10}(b) for the
R\'enyi scattering entropy and Figs. \ref{fig:fig10}(c) and
\ref{fig:fig10}(d) for Tsallis scattering entropy.
In both cases, we obtained the entropies for $C_7$ and $C_8$, where we
observe that for the cycle quantum graphs with an odd number of vertices,
the entropies are zero for the values of $k \ell = \pi + 2j\pi$, with $j
= 1, 2, 3,...$, the same points where the reflection is maximum and all
the other scattering coefficients for any other channel are equal to
zero.
We notice that the scattering entropies are different, so they can be
used to induce distinct results under the same quantum graph type.

\subsection{Wheel quantum graphs with $n$ leads}

The wheel graphs, $W_n$, are defined as a set of $n$ vertices, in which
one vertex has degree $n-1$, which here we call $v_i$, while all the
others have the same degree ($3$), and are labeled as $v_f$.
Thus, to define this quantum graph with $n$ channels in the most
symmetrical way, we consider the central vertex as the entrance channel,
while the others are all exit channels.
So, the vertex $v_i$ now has degree $n$, while all other vertices have
degree $4$.
See Fig. \ref{fig:fig8}(b) for an illustration.

Due to the symmetry and as all the vertices have the same boundary
condition, we get the same transmission amplitude for these $n-1$
vertices.
Therefore, the expression for the reflection and transmission amplitudes
in the wheel quantum graph with Neumann boundary conditions are,
respectively,
\begin{equation}
  \sigma_{W_n}^{\{v_i,v_i\}} =
  \frac
  {4-n\left(2+z^{2}-z^{3}\right)}
  {2 n+(n-2)\left(z^{2}-z^{3}\right)},
\end{equation}
and
\begin{equation}
  \sigma_{n_{W_n}}^{\{v_f,v_i\}} =
  \frac
  {2 z \left(1+z\right)}
  {2 n+(n-2)\left(z^{2}-z^{3}\right)}.
\end{equation}

\begin{figure}[b]
  \centering
  \includegraphics[width=\twodpwidth]{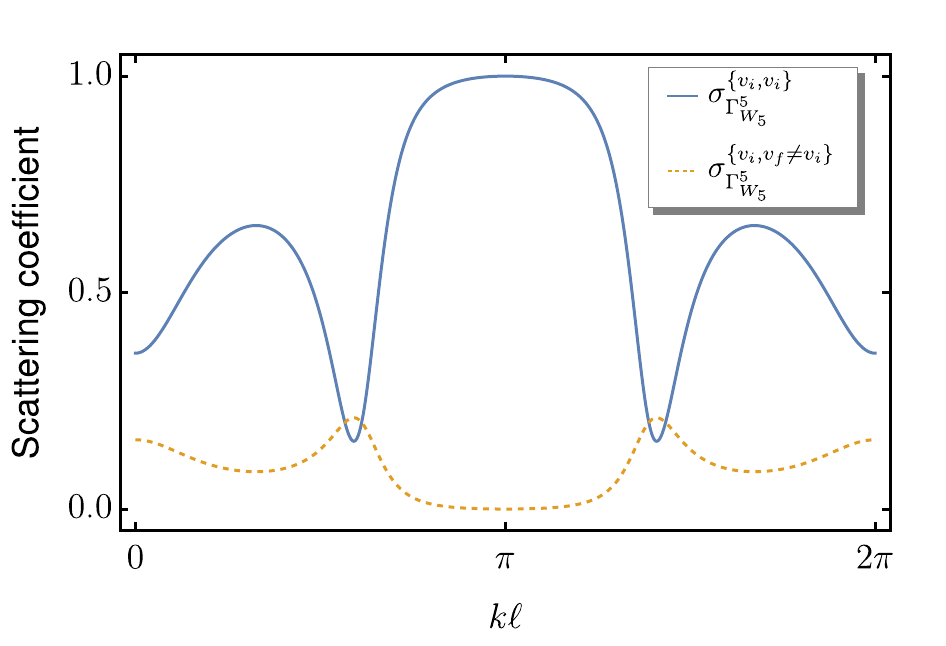}
  \caption{
   The reflection (blue) and transmission (yellow)  coefficients of the
   wheel quantum graph with $5$ vertices.
  }
  \label{fig:fig11}
\end{figure}

\begin{figure}[t]
  \centering
  \includegraphics[width=\threedpwidth]{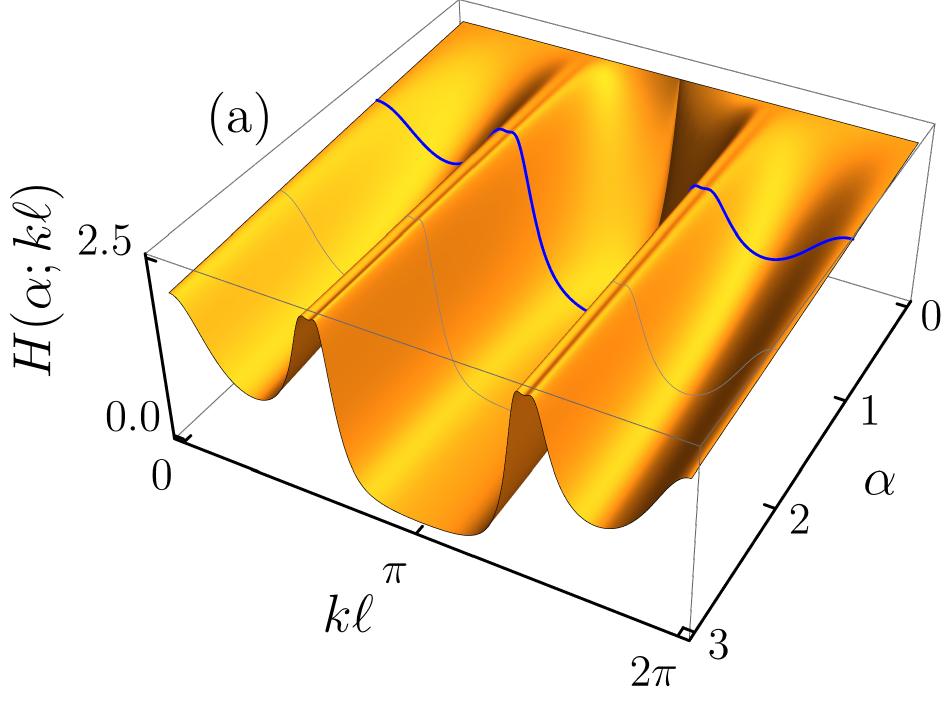}
  \quad
  \includegraphics[width=\threedpwidth]{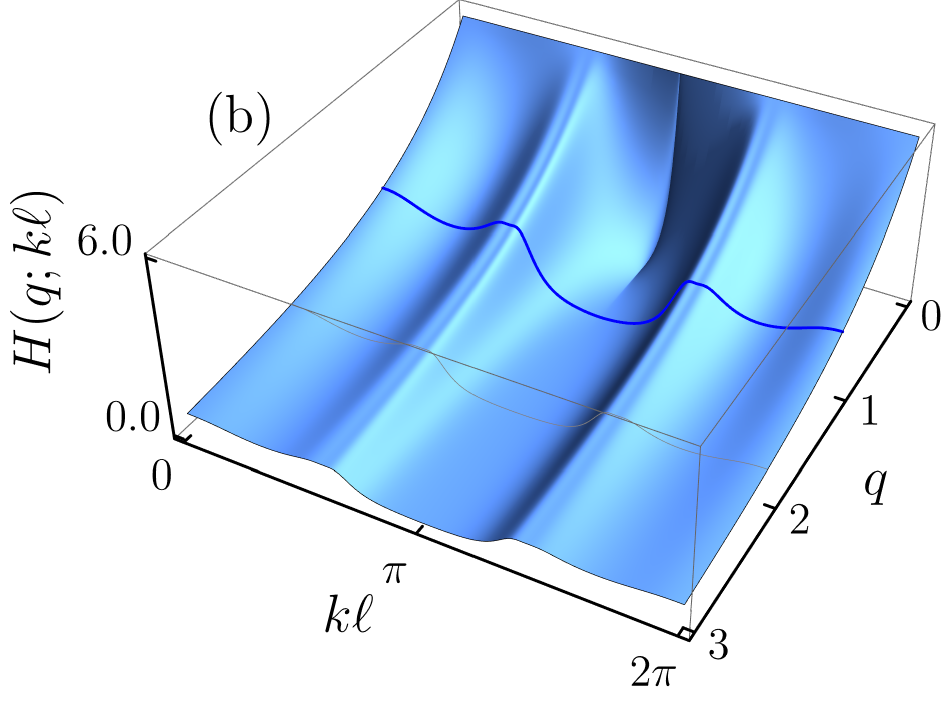}
  \caption{
   (a) The scattering R\'enyi and (b) Tsallis entropies, for the
   wheel quantum graph with $5$ vertices. The limit for the Shannon
   entropy is highlighted in dark blue.
  }
  \label{fig:fig12}
\end{figure}

We display the above results for the scattering entropies in Fig.
\ref{fig:fig11}, in the case of a wheel quantum graph with $5$
vertices.
Also, in Fig. \ref{fig:fig12} we display both the R\'enyi and Tsallis
scattering entropies, highlighting the case of the Shannon entropy in
blue for the same wheel graph with $5$ vertices.
We notice that the scattering entropies are different, so they can be
used to induce distinct results under the same quantum graph type.

\subsection{Complete quantum graphs with $n$ leads}

In complete graphs, $K_n$, we have a set of $n$ vertices, where each
one is connected to all other $n-1$ vertices.
With the addition of the $n$ leads, each vertex has degree $n$.
This is illustrated in Fig. \ref{fig:fig8}(c).
In this case, due to the symmetry of these quantum graphs, any lead
taken as the entrance we have the same expressions for the scattering
amplitudes.
With the Neumann boundary conditions in each vertex, we then get the
reflection amplitude in the form
\begin{equation}
    \sigma_{K_n}^{\{v_i,v_i\}} = \frac{(n-2) \left[(n-4)z-n\left(1+z^{2}-z^{3}\right)
   \right]}{n^2- n (n-4)z+(n-2)^2\left(z^{2}-z^{3}\right) },
\end{equation}
and the transmission amplitude as
\begin{equation}
  \sigma_{K_n}^{\{v_f, v_i\}} =
  \frac{4 z \left(1+z\right)}
  {n^2- n (n-4)z+(n-2)^2\left(z^{2}-z^{3}\right) }.
\end{equation}

\begin{figure}[t]
  \centering
  \includegraphics[width=\twodpwidth]{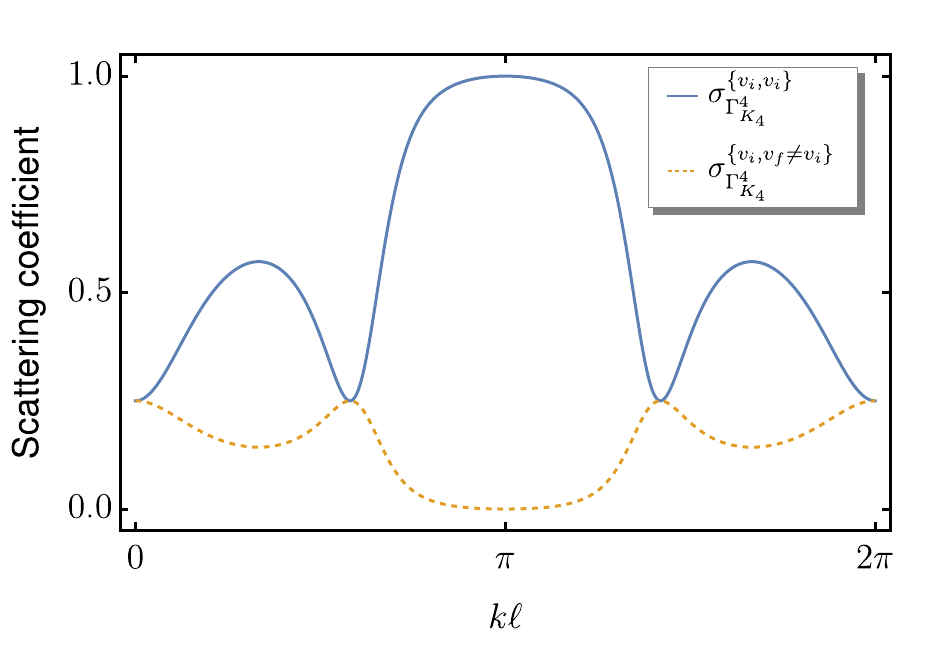}
  \caption{
   The reflection (blue) and transmission (yellow)  coefficients of the
   complete quantum graph with $4$  vertices.
  }
  \label{fig:fig13}
\end{figure}

\begin{figure}[b]
  \centering
  \includegraphics[width=\threedpwidth]{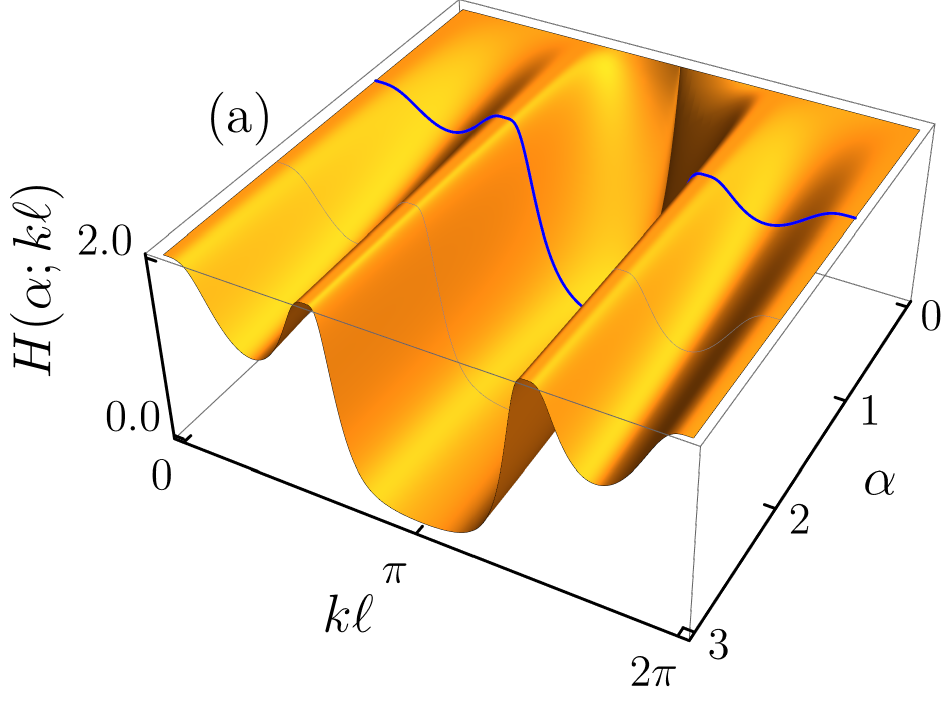}
  \quad
  \includegraphics[width=\threedpwidth]{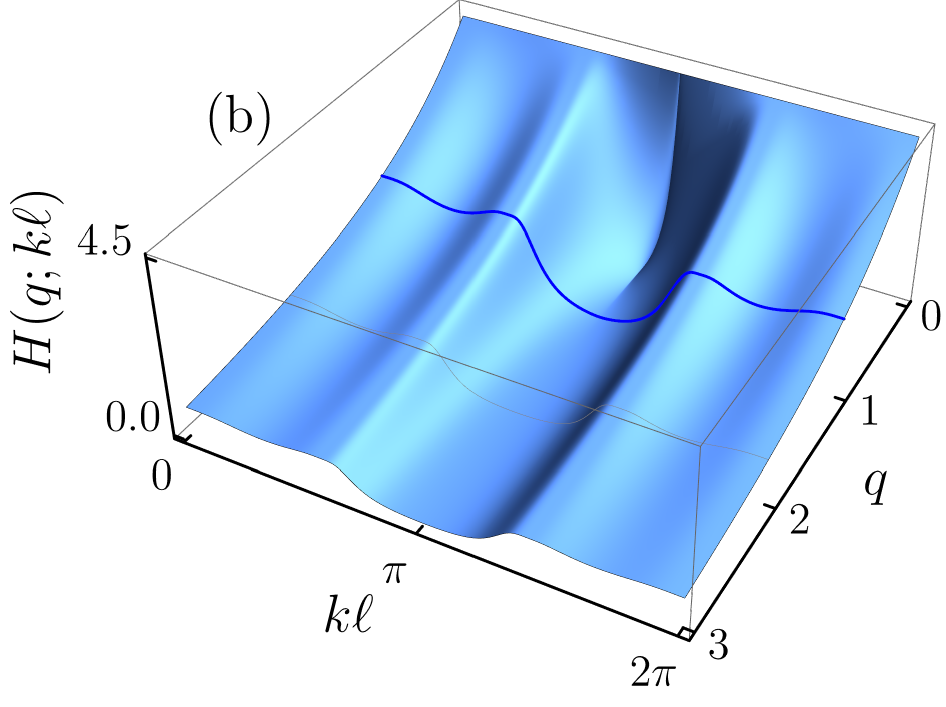}
  \caption{
   (a) The scattering R\'enyi  and (b) Tsallis entropies, for the
   complete quantum graph with $4$ vertices.
   The limit for the Shannon  entropy is highlighted in blue.
  }
  \label{fig:fig14}
\end{figure}

We display the above results for the scattering entropies in
Fig. \ref{fig:fig13}, in the case of a complete graph with $4$
vertices.
Also, in Fig. \ref{fig:fig14} we display both the R\'enyi and Tsallis
scattering entropies, highlighting the case of the Shannon entropy in
blue for the same complete graph with $4$ vertices.
We notice that the scattering entropies are different, so they can be
used to induce distinct results under the same quantum graph type.

\subsection{Average scattering entropies}

\begin{figure}[t]
  \centering
  \includegraphics[width=\twodpwidth]{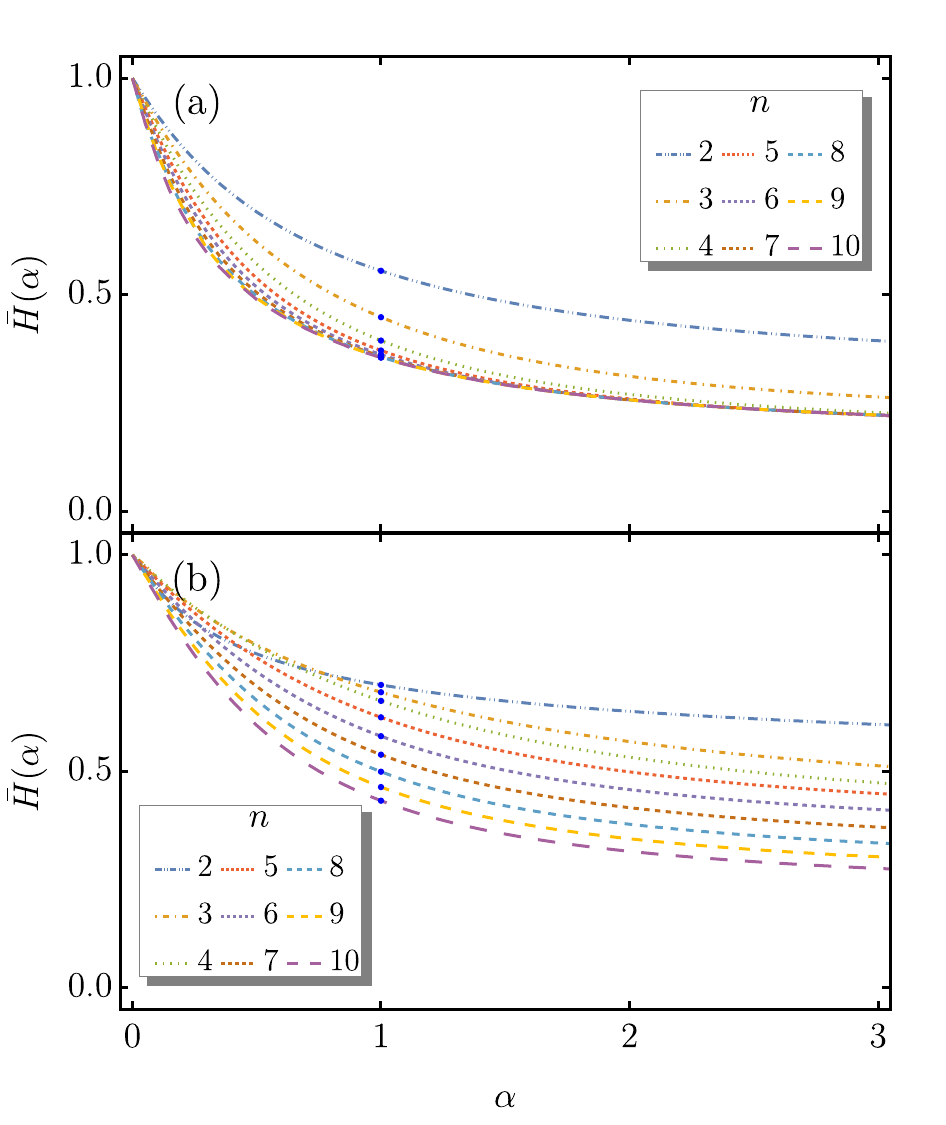}
  \quad
  \includegraphics[width=\twodpwidth]{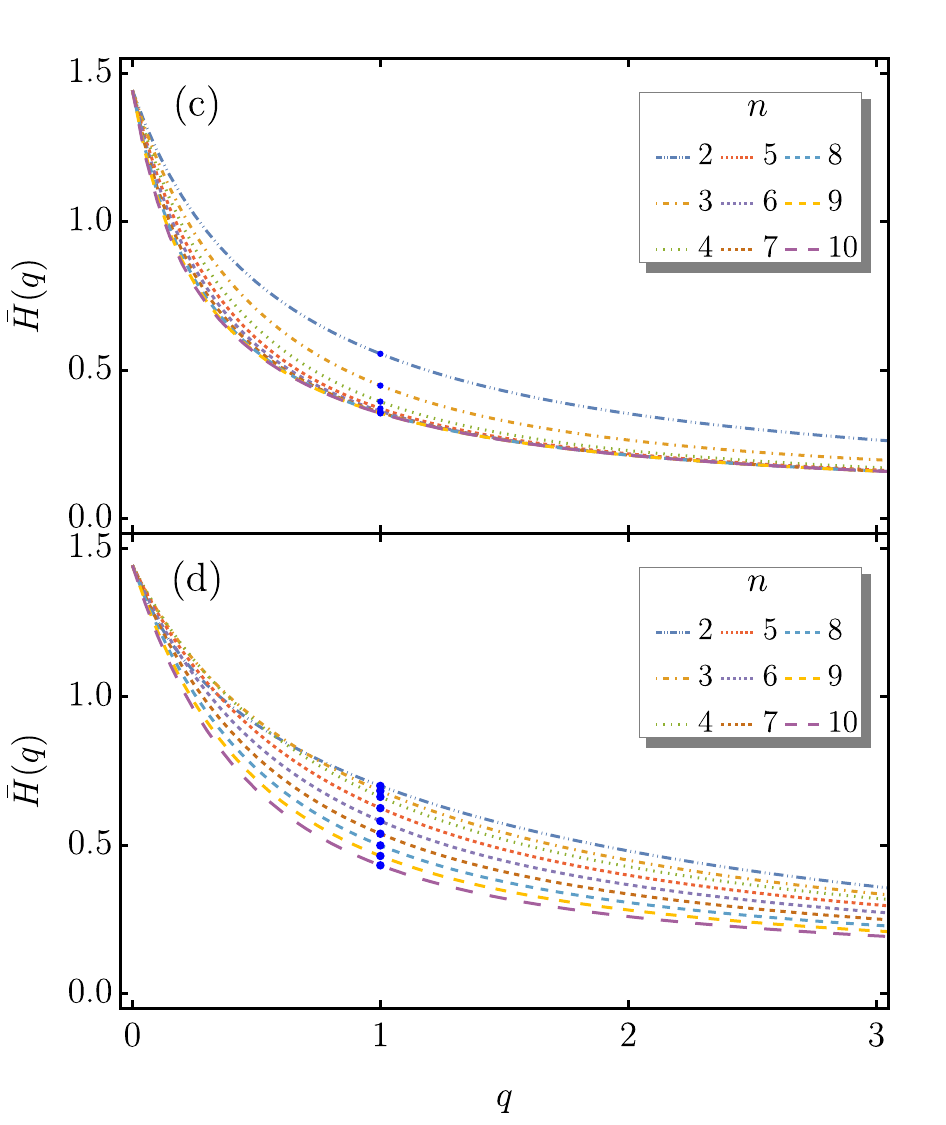}
  \caption{
    Average scattering R\'enyi entropy for (a) series and (b) parallel
    quantum graphs $P(v,v)$ and the average scattering Tsallis entropy
    for (c) series and (d) parallel  quantum graphs $P(v,v)$, for
    several distinct values of $n$.
    The dark blue dots at $q=1$ show the values using the Shannon
    entropy.
  }
  \label{fig:fig15}
\end{figure}

\begin{figure}[t]
  \centering
  \includegraphics[width=\twodpwidth]{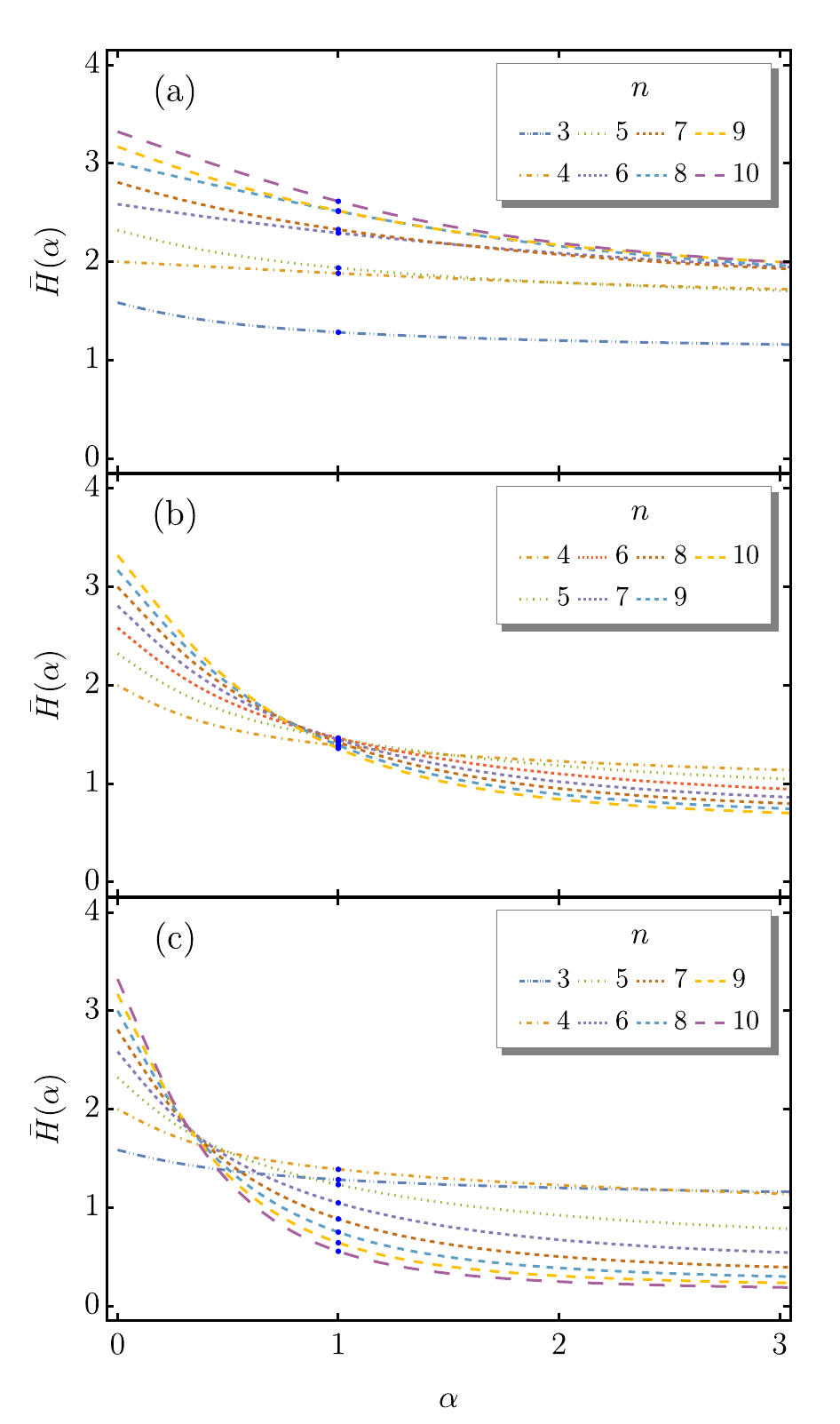}
  \quad
  \includegraphics[width=\twodpwidth]{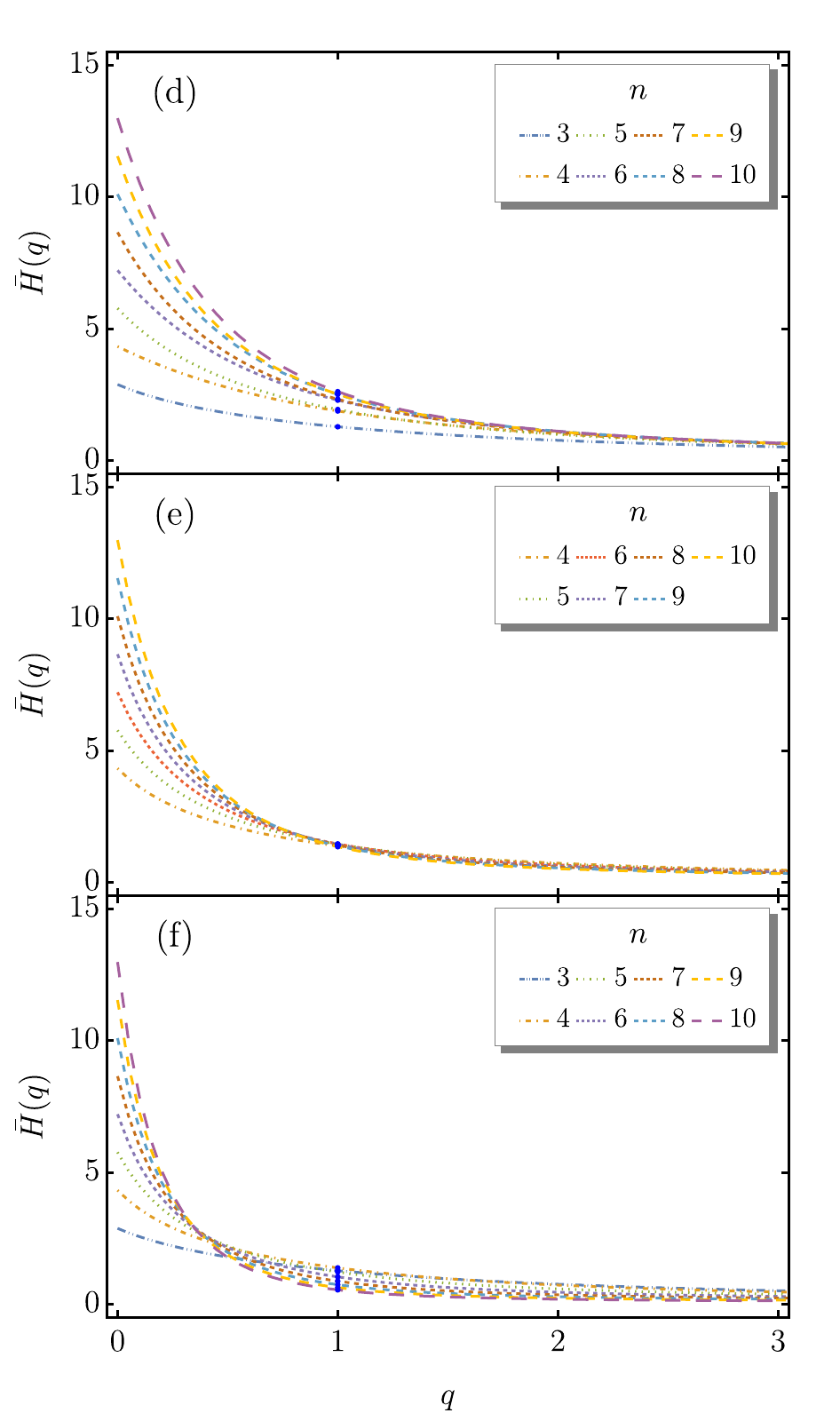}
  \caption{
    Average scattering R\'enyi entropy for the (a) cycle, (b) wheel and
    (c) complete quantum graphs, and
    average scattering Tsallis entropy for the (d) cycle, (e) wheel and
    (f) complete quantum graphs, for several distinct values of $n$.
   The dark blue dots at $\alpha=1$ show the values using the Shannon
   entropy.
 }
  \label{fig:fig16}
\end{figure}

Since all the quantum graphs that we have studied in previous sections
have periodic scattering probabilities, we can also study their average
scattering entropies, which are given in Eq. \eqref{eq:RT-AS} on general
grounds.
We implemented distinct investigations, and in Fig. \ref{fig:fig15}  we
show the results for some series and for parallel arrangements.
The values of the average scattering entropy decrease as one increases
the parameter $\alpha$ for R\'enyi and the parameter $q$ for Tsallis.
Similar results are also depicted in Fig. \ref{fig:fig16}, for distinct
graphs of the cycle, wheel, and complete graphs.
The values of the average scattering R\'enyi and Tsallis entropies also
decrease when one increases $\alpha$ and $q$, respectively.
The results depicted in Figs. \ref{fig:fig15} and \ref{fig:fig16} show
the general behavior, that the average entropies always decrease, as one
increases the parameters $\alpha$ and $q$.

\section{Conclusion}

In this work, we investigated scattering and entropies in distinct types
of quantum graphs.
The process of scattering in quantum graphs is developed by using the
Green's function approach to computing the quantum coefficients in
quantum graphs.
After describing the procedure on general grounds, we added the
scattering entropy and the average scattering entropy using the concept
of entropy due to Shannon, and the respective generalizations to the
case of R\'enyi and Tsallis entropies.
Since we have already considered the case of Shannon in previous works,
the use of the R\'enyi and Tsallis entropies can be implemented
similarly, so they are somehow easy to describe and understand, and can
be manipulated to lead us back to the case of the Shannon entropy within
trustable mathematical manipulations.

After describing the general results, we implemented specific
computations to calculate the scattering R\'enyi and Tsallis entropies
in several distinct situations, described by types of graphs that in
general appear in the study of quantum graphs, among them the series and
parallel, and the circle, wheel and complete graphs.
Since the R\'enyi and Tsallis entropies depend on distinct real
parameters, we developed results with the parameters evaluated within
intervals of current interest, including the cases where they reproduce
the Shannon entropy.
The results show distinct behavior, and the scattering R\'enyi and
Tsallis entropies may induce very different values.
This is interesting since the difference can be identified easily.
The important distinctions among them, depending on the wave number of
the incoming signal may certainly be used in applications, in
particular as a filter to block the quantum transport under certain
circumstances.
The use of quantum graphs as filters was already considered in
\cite{JPSJ.78.124004.2009,EMCE.52599.2021,AP.330.104.2013} and we can
also think of applying the above results in this line of study.

These general results indicate that for all types of quantum graphs
that we studied above, if one thinks of the R\'enyi entropy as an index
of diversity \cite{IS.105.101.1998}, and the Tsallis entropy as an index of nonadditivity \cite{EPL.57.329.2002} in
the same quantum graph, the index of diversity is higher than or equal
to the index of nonadditivity when $\alpha=q\geq1$, but it is lower than
or equal to when $\alpha=q\leq1$.
For the Tsallis entropy, we know that given two independent systems $A$
and $B$, with joint probability density obeying $p(A,B)=p(A)+p(B)$, we
have for the entropy that $H^q(A,B)=H^q(A)+H^q(B)+(1-q)H^q(A)H^q(B)$,
showing that $H^q(A,B)>H^q(A)+H^q(B)$ for $q<1$ and
$H^q(A,B)<H^q(A)+H^q(B)$ for $q>1$.
For the R\'enyi entropy, we know that when $\alpha$ diminishes towards
zero, it increasingly weighs all events with nonzero probability more
equally, regardless of their probabilities.
However, when $\alpha$ increases towards infinity, it is increasingly
determined by the events of the highest probability.
In this sense, for $\alpha$ diminishing towards zero, the R\'enyi
entropy counts an increasing number of accessible states, increasing
both the scattering R\'enyi entropy and the average scattering R\'enyi
entropies.
However, for $\alpha$ increasing to higher and higher values, the
R\'enyi entropy seems to select the events of higher and higher
probabilities, diminishing both the scattering and the average
scattering entropies.
There are other interesting results: for the Tsallis entropy, in
particular, from the average scattering entropy displayed in
Fig. \ref{fig:fig16}(e) for the wheel family of quantum graphs,
one notices that the average Shannon entropy does not significantly
depend on the number of vertices in the quantum graph, but there are
significant variation of the results for $q$ diminishing towards zero,
where the entropy of a joint system seems to be higher than the addition
of the entropies of its individual parts.

The above results are of current interest, and they may be used to
further study specific systems.
In particular, the work \cite{PRE.69.056205.2004}, established an
important connection between quantum graphs and microwave networks, and
this may be further investigated to strengthen the results of the
present work.
For instance, in the recent work \cite{PRE.104.045211.2021}, the authors
described spectral duality in graphs and microwave networks, which can
be further considered in the case of the open quantum graphs studied above.
This is based on the fact that the connection between quantum graphs and
microwave networks work for both spectral and scattering properties.
The results of the present work can also be used in several other
distinct directions, in particular, in the case of reflection and
transmission in ring-like systems with symmetric and asymmetric triple
junctions \cite{JPA.53.155302.2020} and in applications concerning to
the quantum transport in single-molecule junctions that is, in devices
in which a molecule is electrically connected by two electrodes; see,
e.g., the recent reviews \cite{NRP.1.211.2019,NRP.1.381.2019} and
references therein. 
Another line of investigation is related to graph states, as discussed in Refs. \cite{Inproceedings.2006.Hein,PRA.106.012410.2022}.
We think it would be of interest to find some connection between the results of the present work and the entropy studied in these articles.
We hope to report on some related specific issues soon.

\section*{Acknowledgments}
This work was partially supported by the Brazilian agencies Conselho
Nacional de Desenvolvimento Cient\'ifico e Te\-cnol\'ogico (CNPq), Instituto
Nacional de Ci\^{e}ncia e Tecnologia de Informa\c{c}\~{a}o Qu\^{a}ntica
(INCT-IQ), and Para\'iba State Research Foundation (FAPESQ-PB, Grant
0015/2019). It was also financed by the Co\-or\-dena\c{c}\~{a}o de
Aperfei\c{c}oamento de Pessoal de N\'{i}vel Superior (CAPES, Finance
Code 001).
FMA and DB also acknowledge financial support by CNPq Grants 314594/2020-5 (FMA) and
303469/2019-6 (DB).

\section*{Data Availability Statement}
No Data associated in the manuscript.


\begin{thebibliography}{59}
\ifx \bisbn   \undefined \def \bisbn  #1{ISBN #1}\fi
\ifx \binits  \undefined \def \binits#1{#1}\fi
\ifx \bauthor  \undefined \def \bauthor#1{#1}\fi
\ifx \batitle  \undefined \def \batitle#1{#1}\fi
\ifx \bjtitle  \undefined \def \bjtitle#1{#1}\fi
\ifx \bvolume  \undefined \def \bvolume#1{\textbf{#1}}\fi
\ifx \byear  \undefined \def \byear#1{#1}\fi
\ifx \bissue  \undefined \def \bissue#1{#1}\fi
\ifx \bfpage  \undefined \def \bfpage#1{#1}\fi
\ifx \blpage  \undefined \def \blpage #1{#1}\fi
\ifx \burl  \undefined \def \burl#1{\textsf{#1}}\fi
\ifx \doiurl  \undefined \def \doiurl#1{\url{https://doi.org/#1}}\fi
\ifx \betal  \undefined \def \betal{\textit{et al.}}\fi
\ifx \binstitute  \undefined \def \binstitute#1{#1}\fi
\ifx \binstitutionaled  \undefined \def \binstitutionaled#1{#1}\fi
\ifx \bctitle  \undefined \def \bctitle#1{#1}\fi
\ifx \beditor  \undefined \def \beditor#1{#1}\fi
\ifx \bpublisher  \undefined \def \bpublisher#1{#1}\fi
\ifx \bbtitle  \undefined \def \bbtitle#1{#1}\fi
\ifx \bedition  \undefined \def \bedition#1{#1}\fi
\ifx \bseriesno  \undefined \def \bseriesno#1{#1}\fi
\ifx \blocation  \undefined \def \blocation#1{#1}\fi
\ifx \bsertitle  \undefined \def \bsertitle#1{#1}\fi
\ifx \bsnm \undefined \def \bsnm#1{#1}\fi
\ifx \bsuffix \undefined \def \bsuffix#1{#1}\fi
\ifx \bparticle \undefined \def \bparticle#1{#1}\fi
\ifx \barticle \undefined \def \barticle#1{#1}\fi
\bibcommenthead
\ifx \bconfdate \undefined \def \bconfdate #1{#1}\fi
\ifx \botherref \undefined \def \botherref #1{#1}\fi
\ifx \url \undefined \def \url#1{\textsf{#1}}\fi
\ifx \bchapter \undefined \def \bchapter#1{#1}\fi
\ifx \bbook \undefined \def \bbook#1{#1}\fi
\ifx \bcomment \undefined \def \bcomment#1{#1}\fi
\ifx \oauthor \undefined \def \oauthor#1{#1}\fi
\ifx \citeauthoryear \undefined \def \citeauthoryear#1{#1}\fi
\ifx \endbibitem  \undefined \def \endbibitem {}\fi
\ifx \bconflocation  \undefined \def \bconflocation#1{#1}\fi
\ifx \arxivurl  \undefined \def \arxivurl#1{\textsf{#1}}\fi
\csname PreBibitemsHook\endcsname

\bibitem[\protect\citeauthoryear{Berkolaiko and
  Kuchment}{2012}]{Book.Berkolaiko.2012}
\begin{bbook}
\bauthor{\bsnm{Berkolaiko}, \binits{G.}},
\bauthor{\bsnm{Kuchment}, \binits{P.}}:
\bbtitle{Introduction to Quantum Graphs}.
\bsertitle{Mathematical surveys and monographs}.
\bpublisher{American Mathematical Society},
\blocation{Providence}
(\byear{2012})
\end{bbook}
\endbibitem

\bibitem[\protect\citeauthoryear{Shannon}{1948}]{BSTJ.27.379.1948}
\begin{barticle}
\bauthor{\bsnm{Shannon}, \binits{C.E.}}:
\batitle{A mathematical theory of communication}.
\bjtitle{Bell Syst. Tech. J.}
\bvolume{27}(\bissue{3}),
\bfpage{379}--\blpage{423}
(\byear{1948})
\doiurl{10.1002/j.1538-7305.1948.tb01338.x}
\end{barticle}
\endbibitem

\bibitem[\protect\citeauthoryear{R{\'e}nyi and Vekerdi}{1970}]{Book.Renyi.1970}
\begin{bbook}
\bauthor{\bsnm{R{\'e}nyi}, \binits{A.}},
\bauthor{\bsnm{Vekerdi}, \binits{L.}}:
\bbtitle{Probability Theory}.
\bsertitle{Applied mathematics and mechanics},
p. \bfpage{670}.
\bpublisher{North-Holland Publishing Company},
\blocation{Amsterdam}
(\byear{1970})
\end{bbook}
\endbibitem

\bibitem[\protect\citeauthoryear{Dong}{2016}]{PRL.116.251602.2016}
\begin{barticle}
\bauthor{\bsnm{Dong}, \binits{X.}}:
\batitle{Shape dependence of holographic {R}{\'{e}}nyi entropy in conformal
  field theories}.
\bjtitle{Phys. Rev. Lett.}
\bvolume{116}(\bissue{25}),
\bfpage{251602}
(\byear{2016})
\doiurl{10.1103/physrevlett.116.251602}
\end{barticle}
\endbibitem

\bibitem[\protect\citeauthoryear{Gong et~al.}{2021}]{PRL.126.160601.2021}
\begin{barticle}
\bauthor{\bsnm{Gong}, \binits{Z.}},
\bauthor{\bsnm{Piroli}, \binits{L.}},
\bauthor{\bsnm{Cirac}, \binits{J.I.}}:
\batitle{Topological lower bound on quantum chaos by entanglement growth}.
\bjtitle{Phys. Rev. Lett.}
\bvolume{126}(\bissue{16}),
\bfpage{160601}
(\byear{2021})
\doiurl{10.1103/physrevlett.126.160601}
\end{barticle}
\endbibitem

\bibitem[\protect\citeauthoryear{Shi and Kim}{2021}]{PRL.126.141602.2021}
\begin{barticle}
\bauthor{\bsnm{Shi}, \binits{B.}},
\bauthor{\bsnm{Kim}, \binits{I.H.}}:
\batitle{Domain wall topological entanglement entropy}.
\bjtitle{Phys. Rev. Lett.}
\bvolume{126}(\bissue{14}),
\bfpage{141602}
(\byear{2021})
\doiurl{10.1103/physrevlett.126.141602}
\end{barticle}
\endbibitem

\bibitem[\protect\citeauthoryear{Belin et~al.}{2013}]{JHEP.2013.59.2013}
\begin{barticle}
\bauthor{\bsnm{Belin}, \binits{A.}},
\bauthor{\bsnm{Hung}, \binits{L.-Y.}},
\bauthor{\bsnm{Maloney}, \binits{A.}},
\bauthor{\bsnm{Matsuura}, \binits{S.}},
\bauthor{\bsnm{Myers}, \binits{R.C.}},
\bauthor{\bsnm{Sierens}, \binits{T.}}:
\batitle{Holographic charged {R}{\'{e}}nyi entropies}.
\bjtitle{J. High Energy Phys.}
\bvolume{2013}(\bissue{12}),
\bfpage{59}
(\byear{2013})
\doiurl{10.1007/jhep12(2013)059}
\end{barticle}
\endbibitem

\bibitem[\protect\citeauthoryear{Bueno et~al.}{2022}]{PRL.129.021601.2022}
\begin{barticle}
\bauthor{\bsnm{Bueno}, \binits{P.}},
\bauthor{\bsnm{Cano}, \binits{P.A.}},
\bauthor{\bsnm{Murcia}, \binits{{\'{A}}.}},
\bauthor{\bsnm{S{\'{a}}nchez}, \binits{A.R.}}:
\batitle{Universal feature of charged entanglement entropy}.
\bjtitle{Phys. Rev. Lett.}
\bvolume{129}(\bissue{2}),
\bfpage{021601}
(\byear{2022})
\doiurl{10.1103/physrevlett.129.021601}
\end{barticle}
\endbibitem

\bibitem[\protect\citeauthoryear{Tsallis}{1988}]{JSP.52.479.1988}
\begin{barticle}
\bauthor{\bsnm{Tsallis}, \binits{C.}}:
\batitle{Possible generalization of {B}oltzmann-{G}ibbs statistics}.
\bjtitle{J. Stat. Phys.}
\bvolume{52}(\bissue{1-2}),
\bfpage{479}
(\byear{1988})
\doiurl{10.1007/bf01016429}
\end{barticle}
\endbibitem

\bibitem[\protect\citeauthoryear{Tsallis}{2009}]{Book.Tsallis.2009}
\begin{bbook}
\bauthor{\bsnm{Tsallis}, \binits{C.}}:
\bbtitle{Introduction to Nonextensive Statistical Mechanics}.
\bpublisher{Springer},
\blocation{New York}
(\byear{2009}).
\doiurl{10.1007/978-0-387-85359-8}
\end{bbook}
\endbibitem

\bibitem[\protect\citeauthoryear{Wong et~al.}{2015}]{PRD.91.114027.2015}
\begin{barticle}
\bauthor{\bsnm{Wong}, \binits{C.-Y.}},
\bauthor{\bsnm{Wilk}, \binits{G.}},
\bauthor{\bsnm{Cirto}, \binits{L.J.L.}},
\bauthor{\bsnm{Tsallis}, \binits{C.}}:
\batitle{From {QCD}-based hard-scattering to nonextensive statistical
  mechanical descriptions of transverse momentum spectra in high-energy $pp$
  and $p\overline{p}$ collisions}.
\bjtitle{Phys. Rev. D}
\bvolume{91},
\bfpage{114027}
(\byear{2015})
\doiurl{10.1103/PhysRevD.91.114027}
\end{barticle}
\endbibitem

\bibitem[\protect\citeauthoryear{Nojiri et~al.}{2021}]{PRD.104.084030.2021}
\begin{barticle}
\bauthor{\bsnm{Nojiri}, \binits{S.}},
\bauthor{\bsnm{Odintsov}, \binits{S.D.}},
\bauthor{\bsnm{Faraoni}, \binits{V.}}:
\batitle{Area-law versus {R}{\'{e}}nyi and {T}sallis black hole entropies}.
\bjtitle{Phys. Rev. D}
\bvolume{104}(\bissue{8}),
\bfpage{084030}
(\byear{2021})
\doiurl{10.1103/physrevd.104.084030}
\end{barticle}
\endbibitem

\bibitem[\protect\citeauthoryear{Sarwar et~al.}{2022}]{EPJC.82.189.2022}
\begin{barticle}
\bauthor{\bsnm{Sarwar}, \binits{G.}},
\bauthor{\bsnm{Hasanujjaman}, \binits{M.}},
\bauthor{\bsnm{Bhattacharyya}, \binits{T.}},
\bauthor{\bsnm{Rahaman}, \binits{M.}},
\bauthor{\bsnm{Bhattacharyya}, \binits{A.}},
\bauthor{\bsnm{Alam}, \binits{J.-e.}}:
\batitle{Nonlinear waves in a hot, viscous and non-extensive quark-gluon
  plasma}.
\bjtitle{Eur. Phys. J. C}
\bvolume{82}(\bissue{3}),
\bfpage{189}
(\byear{2022})
\doiurl{10.1140/epjc/s10052-022-10122-5}
\end{barticle}
\endbibitem

\bibitem[\protect\citeauthoryear{Silva et~al.}{2021}]{PRA.103.062208.2021}
\begin{barticle}
\bauthor{\bsnm{Silva}, \binits{A.A.}},
\bauthor{\bsnm{Andrade}, \binits{F.M.}},
\bauthor{\bsnm{Bazeia}, \binits{D.}}:
\batitle{Average scattering entropy of quantum graphs}.
\bjtitle{Phys. Rev. A}
\bvolume{103}(\bissue{6}),
\bfpage{062208}
(\byear{2021})
\doiurl{10.1103/physreva.103.062208}
\end{barticle}
\endbibitem

\bibitem[\protect\citeauthoryear{Silva et~al.}{2022}]{PE.141.115217.2022}
\begin{barticle}
\bauthor{\bsnm{Silva}, \binits{A.A.}},
\bauthor{\bsnm{Andrade}, \binits{F.M.}},
\bauthor{\bsnm{Bazeia}, \binits{D.}}:
\batitle{Average scattering entropy for periodic, aperiodic and random
  distribution of vertices in simple quantum graphs}.
\bjtitle{Phys. E}
\bvolume{141},
\bfpage{115217}
(\byear{2022})
\doiurl{10.1016/j.physe.2022.115217}
\end{barticle}
\endbibitem

\bibitem[\protect\citeauthoryear{Mandal et~al.}{2013}]{PRL.111.030602.2013}
\begin{barticle}
\bauthor{\bsnm{Mandal}, \binits{D.}},
\bauthor{\bsnm{Quan}, \binits{H.T.}},
\bauthor{\bsnm{Jarzynski}, \binits{C.}}:
\batitle{Maxwell's refrigerator: An exactly solvable model}.
\bjtitle{Phys. Rev. Lett.}
\bvolume{111}(\bissue{3}),
\bfpage{030602}
(\byear{2013})
\doiurl{10.1103/physrevlett.111.030602}
\end{barticle}
\endbibitem

\bibitem[\protect\citeauthoryear{Parrondo et~al.}{2015}]{NP.11.131.2015}
\begin{barticle}
\bauthor{\bsnm{Parrondo}, \binits{J.M.R.}},
\bauthor{\bsnm{Horowitz}, \binits{J.M.}},
\bauthor{\bsnm{Sagawa}, \binits{T.}}:
\batitle{Thermodynamics of information}.
\bjtitle{Nat. Phys.}
\bvolume{11}(\bissue{2}),
\bfpage{131}--\blpage{139}
(\byear{2015})
\doiurl{10.1038/nphys3230}
\end{barticle}
\endbibitem

\bibitem[\protect\citeauthoryear{Gleiser and
  Stamatopoulos}{2012a}]{PLB.713.304.2012}
\begin{barticle}
\bauthor{\bsnm{Gleiser}, \binits{M.}},
\bauthor{\bsnm{Stamatopoulos}, \binits{N.}}:
\batitle{Entropic measure for localized energy configurations: Kinks, bounces,
  and bubbles}.
\bjtitle{Phys. Lett. B}
\bvolume{713}(\bissue{3}),
\bfpage{304}--\blpage{307}
(\byear{2012})
\doiurl{10.1016/j.physletb.2012.05.064}
\end{barticle}
\endbibitem

\bibitem[\protect\citeauthoryear{Gleiser and
  Stamatopoulos}{2012b}]{PRD.86.045004.2012}
\begin{barticle}
\bauthor{\bsnm{Gleiser}, \binits{M.}},
\bauthor{\bsnm{Stamatopoulos}, \binits{N.}}:
\batitle{Information content of spontaneous symmetry breaking}.
\bjtitle{Phys. Rev. D}
\bvolume{86},
\bfpage{045004}
(\byear{2012})
\doiurl{10.1103/PhysRevD.86.045004}
\end{barticle}
\endbibitem

\bibitem[\protect\citeauthoryear{Bernardini et~al.}{2017}]{PLB.765.81.2017}
\begin{barticle}
\bauthor{\bsnm{Bernardini}, \binits{A.E.}},
\bauthor{\bsnm{Braga}, \binits{N.R.F.}},
\bauthor{\bsnm{Rocha}, \binits{R.}}:
\batitle{Configurational entropy of glueball states}.
\bjtitle{Phys. Lett. B}
\bvolume{765},
\bfpage{81}--\blpage{85}
(\byear{2017})
\doiurl{10.1016/j.physletb.2016.12.007}
\end{barticle}
\endbibitem

\bibitem[\protect\citeauthoryear{Braga and da~Rocha}{2018}]{PLB.776.78.2018}
\begin{barticle}
\bauthor{\bsnm{Braga}, \binits{N.R.F.}},
\bauthor{\bsnm{Rocha}, \binits{R.}}:
\batitle{{AdS}/{QCD} duality and the quarkonia holographic information
  entropy}.
\bjtitle{Phys. Lett. B}
\bvolume{776},
\bfpage{78}--\blpage{83}
(\byear{2018})
\doiurl{10.1016/j.physletb.2017.11.034}
\end{barticle}
\endbibitem

\bibitem[\protect\citeauthoryear{Bazeia et~al.}{2019}]{JMMM.475.734.2019}
\begin{barticle}
\bauthor{\bsnm{Bazeia}, \binits{D.}},
\bauthor{\bsnm{Moreira}, \binits{D.C.}},
\bauthor{\bsnm{Rodrigues}, \binits{E.I.B.}}:
\batitle{Configurational entropy for skyrmion-like magnetic structures}.
\bjtitle{J. Magn. Magn. Mater.}
\bvolume{475},
\bfpage{734}--\blpage{740}
(\byear{2019})
\doiurl{10.1016/j.jmmm.2018.12.033}
\end{barticle}
\endbibitem

\bibitem[\protect\citeauthoryear{Bazeia and
  Rodrigues}{2021}]{PLA.392.127170.2021}
\begin{barticle}
\bauthor{\bsnm{Bazeia}, \binits{D.}},
\bauthor{\bsnm{Rodrigues}, \binits{E.I.B.}}:
\batitle{Configurational entropy of skyrmions and half-skyrmions in planar
  magnetic elements}.
\bjtitle{Phys. Lett. A}
\bvolume{392},
\bfpage{127170}
(\byear{2021})
\doiurl{10.1016/j.physleta.2021.127170}
\end{barticle}
\endbibitem

\bibitem[\protect\citeauthoryear{Braga and
  Junqueira}{2021}]{PLB.820.136485.2021}
\begin{barticle}
\bauthor{\bsnm{Braga}, \binits{N.R.F.}},
\bauthor{\bsnm{Junqueira}, \binits{O.C.}}:
\batitle{Configuration entropy in the soft wall {AdS}/{QCD} model and the
  {Wien} law}.
\bjtitle{Phys. Lett. B}
\bvolume{820},
\bfpage{136485}
(\byear{2021})
\doiurl{10.1016/j.physletb.2021.136485}
\end{barticle}
\endbibitem

\bibitem[\protect\citeauthoryear{da~Rocha}{2021}]{PRD.103.106027.2021}
\begin{barticle}
\bauthor{\bsnm{Rocha}, \binits{R.}}:
\batitle{Information entropy in {AdS}/{QCD}: Mass spectroscopy of isovector
  mesons}.
\bjtitle{Phys. Rev. D}
\bvolume{103}(\bissue{10}),
\bfpage{106027}
(\byear{2021})
\doiurl{10.1103/physrevd.103.106027}
\end{barticle}
\endbibitem

\bibitem[\protect\citeauthoryear{Rigobello et~al.}{2022}]{PRD.104.114501.2021}
\begin{barticle}
\bauthor{\bsnm{Rigobello}, \binits{M.}},
\bauthor{\bsnm{Notarnicola}, \binits{S.}},
\bauthor{\bsnm{Magnifico}, \binits{G.}},
\bauthor{\bsnm{Montangero}, \binits{S.}}:
\batitle{Entanglement generation in $(1+1)d$ {QED} scattering processes}.
\bjtitle{Phys. Rev. D}
\bvolume{104}(\bissue{11}),
\bfpage{114501}
(\byear{2022})
\doiurl{10.1103/physrevd.104.114501}
\end{barticle}
\endbibitem

\bibitem[\protect\citeauthoryear{Karapetyan}{2022}]{EPJP.137.590.2022}
\begin{barticle}
\bauthor{\bsnm{Karapetyan}, \binits{G.}}:
\batitle{Nuclear configurational entropy and high-energy hadron-hadron
  scattering reactions}.
\bjtitle{Eur. Phys. J. Plus}
\bvolume{137}(\bissue{5}),
\bfpage{590}
(\byear{2022})
\doiurl{10.1140/epjp/s13360-022-02736-1}
\end{barticle}
\endbibitem

\bibitem[\protect\citeauthoryear{Karapetyan and
  da~Rocha}{2022}]{EPJP.137.762.2022}
\begin{barticle}
\bauthor{\bsnm{Karapetyan}, \binits{G.}},
\bauthor{\bsnm{Rocha}, \binits{R.}}:
\batitle{Nuclear information entropy, gravitational form factor, and glueballs
  in {AdS}/{QCD}}.
\bjtitle{Eur. Phys. J. Plus}
\bvolume{137}(\bissue{7}),
\bfpage{762}
(\byear{2022})
\doiurl{10.1140/epjp/s13360-022-02952-9}
\end{barticle}
\endbibitem

\bibitem[\protect\citeauthoryear{Barreto and
  da~Rocha}{2022}]{PRD.105.064049.2022}
\begin{barticle}
\bauthor{\bsnm{Barreto}, \binits{W.}},
\bauthor{\bsnm{Rocha}, \binits{R.}}:
\batitle{Differential configurational entropy and the gravitational collapse of
  a kink}.
\bjtitle{Phys. Rev. D}
\bvolume{105},
\bfpage{064049}
(\byear{2022})
\doiurl{10.1103/PhysRevD.105.064049}
\end{barticle}
\endbibitem

\bibitem[\protect\citeauthoryear{Barreto et~al.}{2022}]{arXiv:2207.06367}
\begin{barticle}
\bauthor{\bsnm{Barreto}, \binits{W.}},
\bauthor{\bsnm{Herrera-Aguilar}, \binits{A.}},
\bauthor{\bsnm{Rocha}, \binits{R.}}:
\batitle{Configurational entropy of generalized sine-{G}ordon-type models}.
\bjtitle{arXiv}
(\byear{2022})
\doiurl{10.48550/arXiv.2207.06367}
\end{barticle}
\endbibitem

\bibitem[\protect\citeauthoryear{Kottos and Smilansky}{1997}]{PRL.79.4794.1997}
\begin{barticle}
\bauthor{\bsnm{Kottos}, \binits{T.}},
\bauthor{\bsnm{Smilansky}, \binits{U.}}:
\batitle{Quantum chaos on graphs}.
\bjtitle{Phys. Rev. Lett.}
\bvolume{79},
\bfpage{4794}
(\byear{1997})
\doiurl{10.1103/PhysRevLett.79.4794}
\end{barticle}
\endbibitem

\bibitem[\protect\citeauthoryear{Kottos and Smilansky}{1999}]{AoP.274.76.1999}
\begin{barticle}
\bauthor{\bsnm{Kottos}, \binits{T.}},
\bauthor{\bsnm{Smilansky}, \binits{U.}}:
\batitle{Periodic orbit theory and spectral statistics for quantum graphs}.
\bjtitle{Ann. Phys. (NY)}
\bvolume{274},
\bfpage{76}
(\byear{1999})
\doiurl{10.1006/aphy.1999.5904}
\end{barticle}
\endbibitem

\bibitem[\protect\citeauthoryear{Bl{\"{u}}mel
  et~al.}{2002}]{PRL.88.044101.2002}
\begin{barticle}
\bauthor{\bsnm{Bl{\"{u}}mel}, \binits{R.}},
\bauthor{\bsnm{Dabaghian}, \binits{Y.}},
\bauthor{\bsnm{Jensen}, \binits{R.V.}}:
\batitle{Explicitly solvable cases of one-dimensional quantum chaos}.
\bjtitle{Phys. Rev. Lett.}
\bvolume{88}(\bissue{4}),
\bfpage{044101}
(\byear{2002})
\doiurl{10.1103/physrevlett.88.044101}
\end{barticle}
\endbibitem

\bibitem[\protect\citeauthoryear{Bl\"umel et~al.}{2002}]{PRE.65.046222.2002}
\begin{barticle}
\bauthor{\bsnm{Bl\"umel}, \binits{R.}},
\bauthor{\bsnm{Dabaghian}, \binits{Y.}},
\bauthor{\bsnm{Jensen}, \binits{R.V.}}:
\batitle{Exact, convergent periodic-orbit expansions of individual energy
  eigenvalues of regular quantum graphs}.
\bjtitle{Phys. Rev. E}
\bvolume{65}(\bissue{4}),
\bfpage{046222}
(\byear{2002})
\doiurl{10.1103/PhysRevE.65.046222}
\end{barticle}
\endbibitem

\bibitem[\protect\citeauthoryear{Kottos and Schanz}{2001}]{PE.9.523.2001}
\begin{barticle}
\bauthor{\bsnm{Kottos}, \binits{T.}},
\bauthor{\bsnm{Schanz}, \binits{H.}}:
\batitle{Quantum graphs: a model for quantum chaos}.
\bjtitle{Phys. E}
\bvolume{9}(\bissue{3}),
\bfpage{523}
(\byear{2001})
\doiurl{10.1016/S1386-9477(00)00257-5}
\end{barticle}
\endbibitem

\bibitem[\protect\citeauthoryear{Kaplan}{2001}]{PRE.64.036225.2001}
\begin{barticle}
\bauthor{\bsnm{Kaplan}, \binits{L.}}:
\batitle{Eigenstate structure in graphs and disordered lattices}.
\bjtitle{Phys. Rev. E}
\bvolume{64}(\bissue{3}),
\bfpage{036225}
(\byear{2001})
\doiurl{10.1103/PhysRevE.64.036225}
\end{barticle}
\endbibitem

\bibitem[\protect\citeauthoryear{Mateos et~al.}{2022}]{CSF.156.111817.2022}
\begin{barticle}
\bauthor{\bsnm{Mateos}, \binits{D.M.}},
\bauthor{\bsnm{Morana}, \binits{F.}},
\bauthor{\bsnm{Aimar}, \binits{H.}}:
\batitle{A graph complexity measure based on the spectral analysis of the
  {L}aplace operator}.
\bjtitle{Chaos Solitons Fractals}
\bvolume{156},
\bfpage{111817}
(\byear{2022})
\doiurl{10.1016/j.chaos.2022.111817}
\end{barticle}
\endbibitem

\bibitem[\protect\citeauthoryear{Kempe}{2003}]{CP.44.307.2003}
\begin{barticle}
\bauthor{\bsnm{Kempe}, \binits{J.}}:
\batitle{Quantum random walks: an introductory overview}.
\bjtitle{Contemp. Phys.}
\bvolume{44},
\bfpage{307}
(\byear{2003})
\doiurl{10.1080/00107151031000110776}
\end{barticle}
\endbibitem

\bibitem[\protect\citeauthoryear{Tanner}{2006}]{Incollection.2006.Tanner}
\begin{bchapter}
\bauthor{\bsnm{Tanner}, \binits{G.K.}}:
\bctitle{From quantum graphs to quantum random walks}.
In: \bbtitle{Non-Linear Dynamics and Fundamental Interactions}
vol. \bseriesno{213},
pp. \bfpage{69}--\blpage{87}
(\byear{2006}).
\doiurl{10.1007/1-4020-3949-2_6}
\end{bchapter}
\endbibitem

\bibitem[\protect\citeauthoryear{Diaz-Diaz and
  Estrada}{2022}]{CSF.156.111791.2022}
\begin{barticle}
\bauthor{\bsnm{Diaz-Diaz}, \binits{F.}},
\bauthor{\bsnm{Estrada}, \binits{E.}}:
\batitle{Time and space generalized diffusion equation on graph/networks}.
\bjtitle{Chaos Solitons Fractals}
\bvolume{156},
\bfpage{111791}
(\byear{2022})
\doiurl{10.1016/j.chaos.2022.111791}
\end{barticle}
\endbibitem

\bibitem[\protect\citeauthoryear{Andrade and
  Severini}{2018}]{PRA.98.062107.2018}
\begin{barticle}
\bauthor{\bsnm{Andrade}, \binits{F.M.}},
\bauthor{\bsnm{Severini}, \binits{S.}}:
\batitle{Unitary equivalence between the {G}reen{\textquotesingle}s function
  and {S}chrödinger approaches for quantum graphs}.
\bjtitle{Phys. Rev. A}
\bvolume{98}(\bissue{6}),
\bfpage{062107}
(\byear{2018})
\doiurl{10.1103/physreva.98.062107}
\end{barticle}
\endbibitem

\bibitem[\protect\citeauthoryear{Drinko et~al.}{2019}]{PRA.100.62117.2019}
\begin{barticle}
\bauthor{\bsnm{Drinko}, \binits{A.}},
\bauthor{\bsnm{Andrade}, \binits{F.M.}},
\bauthor{\bsnm{Bazeia}, \binits{D.}}:
\batitle{Narrow peaks of full transmission in simple quantum graphs}.
\bjtitle{Phys. Rev. A}
\bvolume{100}(\bissue{6}),
\bfpage{062117}
(\byear{2019})
\doiurl{10.1103/physreva.100.062117}
\end{barticle}
\endbibitem

\bibitem[\protect\citeauthoryear{Drinko et~al.}{2020}]{EPJP.135.451.2020}
\begin{barticle}
\bauthor{\bsnm{Drinko}, \binits{A.}},
\bauthor{\bsnm{Andrade}, \binits{F.M.}},
\bauthor{\bsnm{Bazeia}, \binits{D.}}:
\batitle{Simple quantum graphs proposal for quantum devices}.
\bjtitle{Eur. Phys. J. Plus}
\bvolume{135}(\bissue{6}),
\bfpage{451}
(\byear{2020})
\doiurl{10.1140/epjp/s13360-020-00459-9}
\end{barticle}
\endbibitem

\bibitem[\protect\citeauthoryear{Andrade et~al.}{2016}]{PR.647.1.2016}
\begin{barticle}
\bauthor{\bsnm{Andrade}, \binits{F.M.}},
\bauthor{\bsnm{Schmidt}, \binits{A.G.M.}},
\bauthor{\bsnm{Vicentini}, \binits{E.}},
\bauthor{\bsnm{Cheng}, \binits{B.K.}},
\bauthor{\bsnm{Luz}, \binits{M.G.E.}}:
\batitle{Green's function approach for quantum graphs: an overview}.
\bjtitle{Phys. Rep.}
\bvolume{647},
\bfpage{1}--\blpage{46}
(\byear{2016})
\doiurl{10.1016/j.physrep.2016.07.001}
\end{barticle}
\endbibitem

\bibitem[\protect\citeauthoryear{Lawrie et~al.}{2023}]{JPA.56.475202.2023}
\begin{barticle}
\bauthor{\bsnm{Lawrie}, \binits{T.}},
\bauthor{\bsnm{Gnutzmann}, \binits{S.}},
\bauthor{\bsnm{Tanner}, \binits{G.}}:
\batitle{Closed form expressions for the green’s function of a quantum
  graph—a scattering approach}.
\bjtitle{J. Phys. A Math. Theor.}
\bvolume{56}(\bissue{47}),
\bfpage{475202}
(\byear{2023})
\doiurl{10.1088/1751-8121/ad03a5}
\end{barticle}
\endbibitem

\bibitem[\protect\citeauthoryear{Diestel}{2010}]{Book.2010.Diestel}
\begin{bbook}
\bauthor{\bsnm{Diestel}, \binits{R.}}:
\bbtitle{Graph Theory},
\bedition{4\textsuperscript{th}} edn.
\bsertitle{Graduate Texts in Mathematics Vol. 173}.
\bpublisher{Springer},
\blocation{New York}
(\byear{2010})
\end{bbook}
\endbibitem

\bibitem[\protect\citeauthoryear{Tanner}{2000}]{JPA.33.3567.2000}
\begin{barticle}
\bauthor{\bsnm{Tanner}, \binits{G.}}:
\batitle{Spectral statistics for unitary transfer matrices of binary graphs}.
\bjtitle{J. Phys. A}
\bvolume{33}(\bissue{18}),
\bfpage{3567}
(\byear{2000})
\doiurl{10.1088/0305-4470/33/18/304}
\end{barticle}
\endbibitem

\bibitem[\protect\citeauthoryear{Ahmed et~al.}{2021}]{EMCE.52599.2021}
\begin{bchapter}
\bauthor{\bsnm{Ahmed}, \binits{M.}},
\bauthor{\bsnm{Gradoni}, \binits{G.}},
\bauthor{\bsnm{Creagh}, \binits{S.}},
\bauthor{\bsnm{Tanner}, \binits{G.}}:
\bctitle{Meta-networks: Reconfigurable cable network topologies for
  interference control}.
In: \bbtitle{2021 IEEE International Joint EMC/SI/PI and EMC Europe Symposium},
pp. \bfpage{520}--\blpage{523}
(\byear{2021}).
\doiurl{10.1109/EMC/SI/PI/EMCEurope52599.2021.9559287}
\end{bchapter}
\endbibitem

\bibitem[\protect\citeauthoryear{Cheon et~al.}{2009}]{JPSJ.78.124004.2009}
\begin{barticle}
\bauthor{\bsnm{Cheon}, \binits{T.}},
\bauthor{\bsnm{Exner}, \binits{P.}},
\bauthor{\bsnm{Turek}, \binits{O.}}:
\batitle{Spectral filtering in quantum y-junction}.
\bjtitle{J. Phys. Soc. Jpn}
\bvolume{78}(\bissue{12}),
\bfpage{124004}
(\byear{2009})
\doiurl{10.1143/JPSJ.78.124004}
\end{barticle}
\endbibitem

\bibitem[\protect\citeauthoryear{Turek and Cheon}{2013}]{AP.330.104.2013}
\begin{barticle}
\bauthor{\bsnm{Turek}, \binits{O.}},
\bauthor{\bsnm{Cheon}, \binits{T.}}:
\batitle{Potential-controlled filtering in quantum star graphs}.
\bjtitle{Ann. Phys. (NY)}
\bvolume{330},
\bfpage{104}--\blpage{141}
(\byear{2013})
\doiurl{10.1016/j.aop.2012.11.011}
\end{barticle}
\endbibitem

\bibitem[\protect\citeauthoryear{Mayoral}{1998}]{IS.105.101.1998}
\begin{barticle}
\bauthor{\bsnm{Mayoral}, \binits{M.M.}}:
\batitle{Renyi's entropy as an index of diversity in simple-stage cluster
  sampling}.
\bjtitle{Information Sciences}
\bvolume{105}(\bissue{1}),
\bfpage{101}--\blpage{114}
(\byear{1998})
\doiurl{10.1016/S0020-0255(97)10025-1}
\end{barticle}
\endbibitem

\bibitem[\protect\citeauthoryear{Beck}{2002}]{EPL.57.329.2002}
\begin{barticle}
\bauthor{\bsnm{Beck}, \binits{C.}}:
\batitle{Non-additivity of tsallis entropies and fluctuations of temperature}.
\bjtitle{Europhysics Letters}
\bvolume{57}(\bissue{3}),
\bfpage{329}
(\byear{2002})
\doiurl{10.1209/epl/i2002-00464-8}
\end{barticle}
\endbibitem

\bibitem[\protect\citeauthoryear{Hul et~al.}{2004}]{PRE.69.056205.2004}
\begin{barticle}
\bauthor{\bsnm{Hul}, \binits{O.}},
\bauthor{\bsnm{Bauch}, \binits{S.}},
\bauthor{\bsnm{Pako\'{n}ski}, \binits{P.}},
\bauthor{\bsnm{Savytskyy}, \binits{N.}},
\bauthor{\bsnm{\.{Z}yczkowski}, \binits{K.}},
\bauthor{\bsnm{Sirko}, \binits{L.}}:
\batitle{Experimental simulation of quantum graphs by microwave networks}.
\bjtitle{Phys. Rev. E}
\bvolume{69}(\bissue{5}),
\bfpage{056205}
(\byear{2004})
\doiurl{10.1103/PhysRevE.69.056205}
\end{barticle}
\endbibitem

\bibitem[\protect\citeauthoryear{Hofmann et~al.}{2021}]{PRE.104.045211.2021}
\begin{barticle}
\bauthor{\bsnm{Hofmann}, \binits{T.}},
\bauthor{\bsnm{Lu}, \binits{J.}},
\bauthor{\bsnm{Kuhl}, \binits{U.}},
\bauthor{\bsnm{St\"ockmann}, \binits{H.-J.}}:
\batitle{Spectral duality in graphs and microwave networks}.
\bjtitle{Phys. Rev. E}
\bvolume{104},
\bfpage{045211}
(\byear{2021})
\doiurl{10.1103/PhysRevE.104.045211}
\end{barticle}
\endbibitem

\bibitem[\protect\citeauthoryear{Fujimoto et~al.}{2020}]{JPA.53.155302.2020}
\begin{barticle}
\bauthor{\bsnm{Fujimoto}, \binits{Y.}},
\bauthor{\bsnm{Konno}, \binits{K.}},
\bauthor{\bsnm{Nagasawa}, \binits{T.}},
\bauthor{\bsnm{Takahashi}, \binits{R.}}:
\batitle{{Quantum Reflection and Transmission in Ring Systems with Double
  Y-Junctions: Occurrence of Perfect Reflection}}.
\bjtitle{J. Phys. A}
\bvolume{53}(\bissue{15}),
\bfpage{155302}
(\byear{2020})
\doiurl{10.1088/1751-8121/ab7601}
\end{barticle}
\endbibitem

\bibitem[\protect\citeauthoryear{Xin et~al.}{2019}]{NRP.1.211.2019}
\begin{barticle}
\bauthor{\bsnm{Xin}, \binits{N.}},
\bauthor{\bsnm{Guan}, \binits{J.}},
\bauthor{\bsnm{Zhou}, \binits{C.}},
\bauthor{\bsnm{Chen}, \binits{X.}},
\bauthor{\bsnm{Gu}, \binits{C.}},
\bauthor{\bsnm{Li}, \binits{Y.}},
\bauthor{\bsnm{Ratner}, \binits{M.A.}},
\bauthor{\bsnm{Nitzan}, \binits{A.}},
\bauthor{\bsnm{Stoddart}, \binits{J.F.}},
\bauthor{\bsnm{Guo}, \binits{X.}}:
\batitle{Concepts in the design and engineering of single-molecule electronic
  devices}.
\bjtitle{Nat. Rev. Phys.}
\bvolume{1}(\bissue{3}),
\bfpage{211}--\blpage{230}
(\byear{2019})
\doiurl{10.1038/s42254-019-0022-x}
\end{barticle}
\endbibitem

\bibitem[\protect\citeauthoryear{Gehring et~al.}{2019}]{NRP.1.381.2019}
\begin{barticle}
\bauthor{\bsnm{Gehring}, \binits{P.}},
\bauthor{\bsnm{Thijssen}, \binits{J.M.}},
\bauthor{\bsnm{Zant}, \binits{H.S.J.}}:
\batitle{Single-molecule quantum-transport phenomena in break junctions}.
\bjtitle{Nat. Rev. Phys.}
\bvolume{1}(\bissue{6}),
\bfpage{381}
(\byear{2019})
\doiurl{10.1038/s42254-019-0055-1}
\end{barticle}
\endbibitem

\bibitem[\protect\citeauthoryear{Hein et~al.}{2006}]{Inproceedings.2006.Hein}
\begin{bchapter}
\bauthor{\bsnm{Hein}, \binits{M.}},
\bauthor{\bsnm{D\"{u}r}, \binits{W.}},
\bauthor{\bsnm{Eisert}, \binits{J.}},
\bauthor{\bsnm{Raussendorf}, \binits{R.}},
\bauthor{\bsnm{Nest}, \binits{M.V.}},
\bauthor{\bsnm{Briegel}, \binits{H.-J.}}:
\bctitle{Entanglement in graph states and its applications}.
In: \beditor{\bsnm{Casati}, \binits{G.}},
\beditor{\bsnm{Shepelyansky}, \binits{D.L.}},
\beditor{\bsnm{Zoller}, \binits{P.}},
\beditor{\bsnm{Benenti}, \binits{G.}} (eds.)
\bbtitle{Proceedings of the International School of Physics "Enrico Fermi"},
pp. \bfpage{115}--\blpage{218}
(\byear{2006}).
\doiurl{10.3254/978-1-61499-018-5-115}
\end{bchapter}
\endbibitem

\bibitem[\protect\citeauthoryear{Zhou and Hamma}{2022}]{PRA.106.012410.2022}
\begin{barticle}
\bauthor{\bsnm{Zhou}, \binits{Y.}},
\bauthor{\bsnm{Hamma}, \binits{A.}}:
\batitle{Entanglement of random hypergraph states}.
\bjtitle{Phys. Rev. A}
\bvolume{106}(\bissue{1}),
\bfpage{012410}
(\byear{2022})
\doiurl{10.1103/physreva.106.012410}
\end{barticle}
\endbibitem

\end{thebibliography}

\end{document}